\documentclass[12pt,thmsb]{article}
\usepackage{amssymb}

\usepackage{amsmath}
\usepackage{harvard}
\usepackage{scalefnt}
\usepackage{lscape}
\usepackage{graphicx}

\renewcommand{\arraystretch}{.5}
\advance \topmargin by -\headheight
\advance \topmargin by -\headsep
\evensidemargin \oddsidemargin
\let \tilde=\widetilde
\newtheorem{theorem}{Theorem}
\newtheorem{Assumption}{Assumption}

\newtheorem{corollary}{Corollary}
\renewcommand{\arraystretch}{1.5}

\input tcilatex
\makeatletter 
\def \@biblabel#1{\hspace*{-\labelsep}}
\makeatother

\begin{document}

\title{Inference Related to Common Breaks in a Multivariate System with
Joined Segmented Trends with Applications to Global and Hemispheric
Temperatures\thanks{%
We are grateful to Tommaso Proietti, Eric Hillebrand and a referee for
useful comments, as well as the participants at the International
Association for Applied Econometrics 2017 Annual Conference. Kim's work was
supported by the Korea University Future Researc Grants (K1720611). Oka's
work was supported by the Singapore Ministry of Education Academic Research
Fund Tier 1 (FY2015-FRC3-003) while Oka was affiliated with the National
University of Singapore and subsequently supported by a Start-up grant from
Monash University.}}
\author{\textbf{Dukpa Kim}\thanks{%
Department of Economics, Korea University, 145 Anam-ro, Seongbuk-gu, Seoul,
02841 Korea (dukpakim@korea.ac.kr).} \\
%EndAName
\textbf{Korea University} \and \textbf{Tatsushi Oka}\thanks{%
Department of Econometrics and Business Statistics, Monash University, 900
Dandenong Road, Caulfield East, Victoria 3145, Australia
(tatsushi.oka@monash.edu).} \\
%EndAName
\textbf{Monash University} \and \textbf{Francisco Estrada}\thanks{%
Centro de Ciencias de la Atm\'{o}sfera, Universidad Nacional Aut\'{o}noma de
M\'{e}xico, Ciudad Universitaria, Circuito Exterior, 0451 M\'{e}xico, DF,
Mexico; and Institute for Environmental Studies, Vrije Universiteit,
Amsterdam, Netherlands (feporrua@atmosfera.unam.mx).} \\
%EndAName
\textbf{Universidad Nacional Aut\'{o}noma de M\'{e}xico and VU University
Amsterdam} \and \textbf{Pierre Perron}\thanks{%
Department of Economics, Boston University, 270 Bay State Rd., Boston, MA,
02215 (perron@bu.edu).} \\
%EndAName
\textbf{Boston University}}
\date{January 28, 2017; Revised: April 4, 2018}
\maketitle

\newpage 

\thispagestyle{empty}

\begin{abstract}
What transpires from recent research is that temperatures and radiative
forcing seem to be characterized by a linear trend with two changes in the
rate of growth. The first occurs in the early 60s and indicates a very large
increase in the rate of growth of both temperature and radiative forcing
series. This was termed as the ``onset of sustained global warming''. The
second is related to the more recent so-called hiatus period, which suggests
that temperatures and total radiative forcing have increased less rapidly
since the mid-90s compared to the larger rate of increase from 1960 to 1990.
There are two issues that remain unresolved. The first is whether the breaks
in the slope of the trend functions of temperatures and radiative forcing
are common. This is important because common breaks coupled with the basic
science of climate change would strongly suggest a causal effect from
anthropogenic factors to temperatures. The second issue relates to
establishing formally via a proper testing procedure that takes into account
the noise in the series, whether there was indeed a `hiatus period' for
temperatures since the mid 90s. This is important because such a test would
counter the widely held view that the hiatus is the product of natural
internal variability. Our paper provides tests related to both issues. The
results show that the breaks in temperatures and radiative forcing are
common and that the hiatus is characterized by a significant decrease in
their rate of growth. The statistical results are of independent interest
and applicable more generally.\vspace{0.1in}

\vspace{0.05in}

\newpage \thispagestyle{empty}
\end{abstract}

\setcounter{page}{1}\baselineskip=18.0pt

\section{Introduction}

Significant advances have been made in documenting how global and
hemispheric temperatures have evolved and in learning about the causes of
these changes. On the one hand, large efforts have been devoted to
investigate the time series properties of temperature and radiative forcing
variables (Gay-Garcia et al., 2009; Kaufmann et al., 2006; Mills, 2013; Tol
and de Vos, 1993). In addition, a variety of methods were applied to detect
and model the trends in climate variables, including features such as breaks
and nonlinearities (Estrada, Perron and Mart\'{\i}nez-L\'{o}pez, 2013;
Gallagher et al., 2013; Harvey and Mills, 2002; Karl et al., 2000; Pretis et
al., 2015; Reeves et al., 2007; Seidel and Lanzante, 2004; Stocker et al.,
2013; Tom\'{e} and Miranda, 2004). Multivariate models of temperature and
radiative forcing series provide strong evidence for a common secular trend
between these variables, and help to evaluate the relative importance of its
natural and anthropogenic drivers (Estrada, Perron and Mart\'{\i}nez-L\'{o}%
pez, 2013; Estrada, Perron, Gay-Garc\'{\i}a and Mart\'{\i}nez-L\'{o}pez,
2013; Kaufmann et al., 2006; Tol and Vos, 1998). The methodological
contributions of the econometrics literature to this field have been
notable; e.g., Dickey and Fuller (1979), Engle and Granger (1987), Johansen
(1991), Perron (1989, 1997), Bierens (2000), Ng and Perron (2001), Kim and
Perron (2009), Perron and Yabu (2009), among many others, see Estrada and
Perron (2014) for a review. Regardless of the differences in assumptions and
methods (statistical- or physical), there is a general consensus about the
existence of a common secular trend between temperatures and radiative
forcing variables.

Some of the most relevant questions about the attribution of climate change
are concerned with understanding particular periods in which the rate of
warming changed; e.g., rapid warming, slowdowns or pauses. An example is the
apparent slowdown in warming since the 1990s, for which various methods were
applied to document or reject its existence (e.g., Fyfe et al., 2016;
Lewandowsky et al., 2015, 2016). These issues are delicate to handle when
dealing with observed temperature series, as the object of interest is the
secular trend behind the observed warming. This underlying trend is affected
by low-frequency natural variability (Swanson et al., 2009; Wu et al., 2011)
and changes in its rate of warming are difficult to detect and attribute.
New tests and approaches to investigate common features in temperature and
radiative forcing can make attribution studies more relevant for climate
science and policy-making by providing a better understanding of the drivers
behind them.

One way to tackle this problem is to devise procedures to extract the common
secular trend between temperature and radiative forcing series and framing
the problem in a univariate context where the available structural change
tests can be applied. Estrada and Perron (2016) used this approach to
investigate the existence and causes of the current slowdown in the warming.
Their results strongly suggests a common secular trend and common breaks,
largely determined by the anthropogenic radiative forcing. Their analysis is
based on the results of co-trending and principal component analyses to
separate the common long-term trend imparted by radiative forcing from the
natural variability component in global temperature series. As they
discussed, filtering the effects of physical modes of natural variability
from temperature series is necessary to obtain a proper assessment of the
features and drivers of the warming trend. This problem has seldom been
addressed within the time-series based attribution literature (e.g.,
Estrada, Perron and Mart\'{\i}nez-L\'{o}pez, 2013; Estrada and Perron, 2016)
and it constitutes a relevant research topic that requires the development
of new procedures and techniques. What transpires from this research is that
temperatures and radiative forcing are most likely characterized by a linear
trend with two changes in the rate of growth. The first occurs in the early
60s and indicates a very large increase in the rate of growth of both
temperatures and radiative forcing. This was termed as the ``onset of
sustained global warming''. The second is related to the more recent
so-called ``hiatus'' period, which suggests that temperatures and total
radiative forcing have increased less rapidly since the mid-90s compared to
the larger rate of increase from 1960 to 1990.

The approach we take consists in designing statistical tests in a
multivariate setting with joint-segmented trends to investigate the
existence of common breaks and the ``hiatus''. There are two issues that
until now remain unresolved, which this approach can address. The first is
whether the breaks in the slope of the trend functions of temperatures and
radiative forcing are common. This is important because common breaks
coupled with the basic science of climate change would strongly suggest a
causal effect from anthropogenic factors to temperatures. As is well known,
a common linear trend can occur when the series are spuriously correlated. A
common break can eliminate such concerns of spurious correlation and foster
the claim that the relationship from radiative forcing to temperatures is
causal. Hence, our paper first aims at developing a test for common breaks
across a set of series modeled as joint segmented trends with correlated
noise. The theoretical framework follows Qu and Perron (2007) and Oka and
Perron (2016), building on Bai and Perron (1998, 2003). Their framework,
however, precludes joint segmented trends since the regressors are functions
of the break dates, which makes the problem very different. Perron and Zhu
(2005) show that the limit distribution of the estimate of the break date in
a single time series with a joint segmented trend follows a normal limit
distribution. We build on that result to show that our common break test
follows a chi-square distribution. Theoretically, our results are more
general and cover a wide range of cases to test hypotheses on the break
dates in a multivariate system, within or across equations. Although our
model with a single equation is similar to that in Perron an Zhu (2005), our
work has a different focus, namely testing, while Perron and Zhu (2005)
deals with the distribution theory of the estimate. We are not aware of any
test in the literature that can handle a set of general linear restrictions
on break dates both across and within equations. Furthermore, our test for
an additional break is more closely related to Perron and Yabu (2009), and
the studies cited therein. The attempt to increase the power of structural
break tests using cross equation correlation is also new in the case of
joint-segmented trends. Our results show that, once we filter temperatures
for the effect of the Atlantic Mutidecadal Oscillation (AMO) and the North
Atlantic Oscillation (NAO) for reasons explained in the text, the breaks in
the slope of radiative forcing and temperatures are common, both for the
large increase in the 60s and the ``hiatus'' period.

The second issue relates to establishing formally via a proper testing
procedure that takes into account the noise in the series, whether there was
a ``hiatus'' period for temperatures since the mid-90s. This is important
because such a test would counter the widely held view that the ``hiatus''
is the product of natural internal variability (Kosaka and Xie, 2013;
Trenberth and Fasullo, 2013; Meehl et al., 2011; Balmaseda, Trenberth and K%
\"{a}ll\'{e}n, 2013). Using standard univariate tests (e.g., Perron and
Yabu, 2009), the results are mixed across various series and sometimes
borderline. Our aim is to provide tests with enhanced power by casting the
testing problem in a bivariate framework involving temperatures and
radiative forcing. Our results indicate that indeed the ``hiatus''\
represents a significant slowdown in the rate of increase in temperatures,
especially when considering global or southern hemisphere series, for which
our test points to a rejection of the null of no change for all data sources.

We consider a multivariate system with $n$ equations where the dependent
variables are modeled as joint-segmented trends with multiple changes in the
slope. The errors are allowed to be serially correlated and correlated
across equations. Of interest is testing for general linear restrictions on
the break dates including testing for common breaks across equations. The
test used is a (quasi-) likelihood ratio test assuming serially uncorrelated
errors, in which case it has a pivotal chi-square distribution. However, it
is non-pivotal in the general case of interest. Accordingly, we also
consider a corrected Wald test with a pivotal limit distribution, which can
be constructed using break dates estimated one equation at a time, labelled
OLS-Wald, or estimated via the complete system, labelled GLS-Wald. The
latter can offer more efficient estimates when there is correlation in the
errors across equations. It is, however, computationally demanding as it
requires least squares operations of order $O(T^{m})$ where $T$ is the
sample size and $m$ is the total number of breaks. The OLS-Wald test
requires only least-squares operations of order $O(T^{m_{1}}+...+T^{m_{n}})$
where $m_{i}$ ($i=1,...,n$) is the number of breaks in the $i^{th}$
equation, making it much easier to compute in a multivariate system. Our
simulation study shows that the two Wald tests have finite sample sizes
close to the nominal size when the extent of the serial correlation in the
noise is small. They, however, suffer from potentially severe liberal size
distortions for moderate to strong serial correlation. Hence, for all three
tests (since the LR\ is non-pivotal), we suggest a bootstrap procedure to
obtain the relevant critical values. These bootstrap tests are shown to have
exact sizes close to the nominal 5\% level in all cases considered.

We also propose a test for the presence of an additional break in some
series. For instance, in our applications, the null and alternative
hypotheses specify a break in the slope of the radiative forcing series (for
which a rejection is easy using a single equation test), while under the
null hypothesis there is no break in the temperature series but there is one
under the alternative. Hence, the aim is to see whether a break in
temperatures is present using a bivariate system. The theoretical framework
is more general and allows multiple breaks in a general multivariate system.
The test considered is a quasi-likelihood ratio test, whose limit
distribution is non-pivotal and depends in a complex way on a number of
nuisance parameters. Hence, we resort to a bootstrap procedure to obtain the
relevant critical values.

The paper is structured as follows. Section 2 presents the statistical model
and the assumptions imposed. Section 3 considers testing for linear
restrictions on the break dates including as a special case testing for
common breaks across equations. The various tests and their limit
distributions are discussed. Section 4 presents the test for the presence of
an additional break in some series within a multivariate system. Section 5
presents simulation results about the size and power of the tests and
introduces the bootstrap versions recommended. Section 6 presents the
applications. The data are discussed in Section 6.1. Section 6.2 presents
the results for the common breaks tests, while Section 6.3 addresses the
issue of testing for the ``hiatus'' using a bivariate system with a
temperature and a radiative forcing series. Brief conclusions are offered in
Section 7. All proofs are in an appendix.

\section{Statistical Model and Assumptions}

We consider the following model with multiple breaks in a system of $n$
variables. Each variable consists of a linear trend with multiple changes in
slope with the trend function is joined at each break date. More
specifically, there are $m_{i}$ breaks in the slope of the $i^{th}$
variable, and the $n$-variate system is:%
\begin{equation}
y_{it}=\mu _{i}+\beta _{i}t+\tsum\nolimits_{j=1}^{m_{i}}\delta
_{ij}b_{t}(k_{ij}^{0})+u_{it},  \label{Model it}
\end{equation}%
for $t=1,...,T$ and $i=1,...,n$, where $b_{t}(k_{ij}^{0})=1(t\geqslant
k_{ij}^{0})(t-k_{ij}^{0})$, $1(A)$ is the indicator function of the event $A$
and\ $k_{ij}^{0}$ is the $j^{th}$ break date for the change (with magnitudes 
$\delta _{ij}$) in the trend of the $i^{th}$ variable. The vector of break
dates for the $i^{th}$ variable is $k_{i}^{0}=(k_{i1}^{0},\dots
,k_{im_{i}}^{0})^{\prime }$ and $k^{0}$ is the vector of break dates for the
entire system, $k^{0}=(k_{1}^{0\prime },\dots ,k_{n}^{0\prime })^{\prime }$.
We let $m=\tsum_{i=1}^{n}m_{i}$ denote the total number of breaks in the
system. Hence, $k^{0}$ is an $m\times 1$ vector. For the $i^{th}$ variable,
we have in matrix form%
\begin{equation}
y_{i}=X(k_{i}^{0})\theta _{i}+u_{i},  \label{Model i}
\end{equation}%
where $X(k_{i}^{0})=[c,\tau ,b(k_{i1}^{0}),\dots ,b(k_{im_{i}}^{0})]$ and $%
\theta _{i}=(\mu _{i},\beta _{i},\delta _{i1},\dots ,\delta
_{im_{i}})^{\prime }$ with $c=(1,\dots ,1)^{\prime }$, $\tau =(1,\dots
,T)^{\prime }$ and $b(k_{ij}^{0})=(b_{1}(k_{ij}^{0}),\dots
,b_{T}(k_{ij}^{0}))^{\prime }$. The entire system in matrix notation is%
\begin{equation}
y=X^{0}\theta +u,  \label{Model}
\end{equation}%
where $y=(y_{1},...,y_{n})^{\prime }$, $%
X^{0}=diag(X(k_{1}^{0}),...,X(k_{n}^{0}))$, $\theta =(\theta _{1}^{\prime
},...,\theta _{n}^{\prime })^{\prime }$ and $u=(u_{1}^{\prime
},...,u_{n}^{\prime })^{\prime }$. We are interested in the case where the
number of breaks for each variable is known but the break dates are unknown.
For a vector of generic break dates, we will use the notation without a $0$
superscript, that is, $k$ and $k_{i}$. Also, we will simply write $X$ to
denote the regressor matrix corresponding to break dates specified by $k$,
that is, $X=diag(X(k_{1}),\dots ,X(k_{n}))$.

To motivate our quasi-likelihood ratio test, we first assume that $u$ is
multivariate normal with 0 mean and covariance $\Sigma \otimes I_{T}$. This
assumption will be relaxed when we derive the asymptotic distribution of our
test. For a generic break date vector $k$ and the corresponding regressor
matrix $X$, the log-likelihood function of $y$ is then given by%
\begin{equation*}
l(k,\theta ,\Sigma )=-(nT/2)\log 2\pi -(T/2)\log \left| \Sigma \right|
-(1/2)(y-X\theta )^{\prime }(\Sigma ^{-1}\otimes I_{T})(y-X\theta )\text{.}
\end{equation*}%
Let $\hat{\theta}=(\hat{\theta}_{1}^{\prime },\dots ,\hat{\theta}%
_{n}^{\prime })^{\prime }$ and $\hat{\Sigma}(k)$ be the maximum likelihood
estimators for $\theta $ and $\Sigma $, which jointly solve $\hat{\theta}%
=[X^{\prime }(\hat{\Sigma}^{-1}(k)\otimes I_{T})X]^{-1}[X^{\prime }(\hat{%
\Sigma}^{-1}(k)\otimes I_{T})y]$ and $\hat{\Sigma}(k)=T^{-1}\hat{U}%
_{k}^{\prime }\hat{U}_{k}$, where $\hat{U}_{k}=[y_{1}-X(k_{1})\hat{\theta}%
_{1},\dots ,y_{n}-X(k_{n})\hat{\theta}_{n}]$. Thus, the maximized the
log-likelihood function is%
\begin{equation}
l(k)=-(nT/2)(\log 2\pi +1)-(T/2)\log |\hat{\Sigma}(k)|\text{.}
\label{loglike}
\end{equation}%
It is useful to work with break fractions; hence define $\lambda
_{ij}=k_{ij}/T$, $\lambda _{i}=k_{i}/T$ and $\lambda =k/T$, with the true
break fractions $\lambda _{ij}^{0}$, $\lambda _{i}^{0}$ and $\lambda ^{0}$
equivalently defined. With these definitions, functions of $\lambda $ such
as $l(\lambda )$ and $\hat{\Sigma}(\lambda )$ will be used interchangeably
with $l(k)$ and $\hat{\Sigma}(k)$. We now state the assumptions needed for
the asymptotic analysis.

\begin{Assumption}
$0<\lambda _{i1}^{0}$ $<...<\lambda _{im_{i}}^{0}<1,$ with $%
k_{ij}^{0}=[T\lambda _{ij}^{0}]$ for $i=1,...,n$.\label{A1}
\end{Assumption}

\begin{Assumption}
$\delta _{ij}\neq 0$ for $j=1,...,m_{i}$ and $i=1,\dots ,n$.\label{A2}
\end{Assumption}

\begin{Assumption}
Let $u_{t}=(u_{1t},\dots ,u_{nt})^{\prime }$. Then, $u_{t}$ is stationary
with $E(u_{t})=0$ and $Var(u_{t})=\Sigma $. In addition, $%
T^{-1/2}\sum_{t=1}^{[Tr]}u_{t}=T^{-1/2}\Psi
^{1/2}\sum_{t=1}^{[Tr]}e_{t}+o_{p}(1)$ and $T^{-1/2}\sum_{t=1}^{[Tr]}e_{t}%
\Rightarrow W(r)$, where ``$\Rightarrow $''\ denotes weak convergence under
the Skorohod topology, $W(r)$ is the $n$ dimensional standard Wiener process
and $\Psi =\lim T^{-1}E(\sum_{t=1}^{T}u_{t})(\sum_{t=1}^{T}u_{t})^{\prime }$.%
\label{A3}
\end{Assumption}

Assumptions \ref{A1} and \ref{A2} are standard and simply state that the
break dates are asymptotically distinct (i.e., each regime increases
proportionally with the sample size $T$) and the changes in the parameters
are non-zero at the break dates. Assumption \ref{A3} states that $u_{t}$ is
a stationary process and its partial sum follows a functional central limit
theorem.

\section{Testing for Linear Restrictions on the Break Dates}

We now consider the following null and alternative hypotheses:%
\begin{equation}
H_{0}:R\lambda ^{0}=r\ \ \ \ \mathrm{and}\ \ \ \ H_{1}:R\lambda ^{0}\neq r%
\text{.}  \label{H0}
\end{equation}%
Some examples of this type of hypotheses are the following: 1) specific
break fractions: $R=I_{m}$ and $r=\bar{\lambda}$; 2) break fractions with a
fixed distance: $R=[0,\dots ,1,0,\dots ,-1,0,\dots ]$ and $r=c>0$; 3) common
breaks: $R=[0,\dots ,1,0,\dots ,-1,0,\dots ]$ and $r=0$. For the first case,
historic events can be used to determine the value of $\bar{\lambda}$. The
second case can be used across equations as well as within an equation. The
third case can be viewed as a special case of the second one and should be
used across equations only. The LR test statistic is%
\begin{equation*}
LR=-2[\max_{\lambda :R\lambda =r}l(\lambda )-\max_{\lambda }l(\lambda
)]=T[\min_{\lambda :R\lambda =r}\log |\hat{\Sigma}(\lambda )|-\min_{\lambda
}\log |\hat{\Sigma}(\lambda )|].
\end{equation*}

\begin{theorem}
Let Assumptions \ref{A1}-\ref{A3} hold with $\Sigma =\Psi $. Then, as $%
T\rightarrow \infty $, we have under $H_{0}$ in (\ref{H0}) that $LR\overset{d%
}{\rightarrow }\chi _{q}^{2}$, where $q=rank(R)$.\label{Theorem 1}
\end{theorem}

Our model with broken segmented trends at unknown dates is a rare case in
structural change problems with the estimates of the break dates having an
asymptotic normal distribution; see Perron and Zhu (2005). Since the
likelihood ratio test essentially involves quadratic forms of these
estimates, it has the usual $\chi ^{2}$ distribution. When the short-run and
long-run variances differ $\Sigma \neq \Psi $, the asymptotic distribution
depends on nuisance parameters. It is difficult to directly modify the LR
test. Instead, we develop a Wald test for the hypotheses (\ref{H0}) robust
to serial correlation in the errors. To express the limiting distribution,
define $f_{i}(r)=(1,r,(r-\lambda _{i1}^{0})^{+},\dots ,(r-\lambda
_{im_{i}}^{0})^{+})^{\prime }$ and $g_{i}(r)=(1(r>\lambda _{i1}^{0}),\dots
,1(r>\lambda _{im_{i}}^{0}))^{\prime }$, with $(\cdot )^{+}$ being the
positive part of the argument. Also, $FF_{ij}=\int f_{i}(r)f_{j}(r)^{\prime
}dr$, $FG_{ij}=\int f_{i}(r)g_{j}(r)^{\prime }dr$, $GF_{ij}=\int
g_{i}(r)f_{j}(r)^{\prime }dr$ and $GG_{ij}=\int g_{i}(r)g_{j}(r)^{\prime }dr$%
, with all integrals from 0 to 1. Denote by $s_{ij}$ the $(i,j)^{th}$
element of $\Sigma ^{-1}$ and by $\kappa _{ij}$ that of $\Sigma ^{-1}\Psi
\Sigma ^{-1}$. Now, define%
\begin{equation*}
Q_{FF}=%
\begin{bmatrix}
s_{11}FF_{11} & \dots & s_{1n}FF_{1n} \\ 
\dots & \ddots & \dots \\ 
s_{n1}FF_{n1} & \dots & s_{nn}FF_{nn}%
\end{bmatrix}%
\text{ and }\Gamma _{FF}=%
\begin{bmatrix}
\kappa _{11}FF_{11} & \dots & \kappa _{1n}FF_{1n} \\ 
\dots & \ddots & \dots \\ 
\kappa _{n1}FF_{n1} & \dots & \kappa _{nn}FF_{nn}%
\end{bmatrix}%
\text{.}
\end{equation*}%
We also define $Q_{FG}$, $Q_{GF}$, $Q_{GG}$, $\Gamma _{FG}$, $\Gamma _{GF}$
and $\Gamma _{GG}$ equivalently, as well as%
\begin{eqnarray*}
\Xi _{0} &=&D_{\delta }\left[ \Gamma _{GG}-\Gamma
_{GF}Q_{FF}^{-1}Q_{FG}-Q_{GF}Q_{FF}^{-1}\Gamma _{FG}+Q_{GF}Q_{FF}^{-1}\Gamma
_{FF}Q_{FF}^{-1}Q_{FG}\right] D_{\delta } \\
\Xi _{1} &=&D_{\delta }\left[ Q_{GG}-Q_{GF}Q_{FF}^{-1}Q_{FG}\right]
D_{\delta }\text{.}
\end{eqnarray*}%
where $D_{\delta }=diag(D_{\delta _{1}},\dots ,D_{\delta _{n}})$ with $%
D_{\delta _{i}}=diag(\delta _{i})$ and $\delta _{i}=(\delta _{i1},\dots
,\delta _{im_{i}})^{\prime }$. We now provide the limiting distribution of
the estimate of the break fraction that maximizes the log-likelihood (\ref%
{loglike}), which allows us to obtain the limit distribution of the Wald
test.

\begin{theorem}
Let Assumptions \ref{A1}-\ref{A3} hold, $\hat{\lambda}=\hat{k}/T$ and $\hat{k%
}=\arg \min_{k}l(k)$. Then, as $T\rightarrow \infty $, (i)\label{Theorem 2} $%
T^{3/2}(\hat{\lambda}-\lambda ^{0})\overset{d}{\rightarrow }N(0,\Xi
_{1}^{-1}\Xi _{0}\Xi _{1}^{-1\prime })$; (ii) under $H_{0}$ in (\ref{H0}),
the Wald test statistic is $Wald\equiv T^{3}(R\hat{\lambda}-r)^{\prime }(R%
\hat{\Xi}R^{\prime })^{-1}(R\hat{\lambda}-r)\overset{d}{\rightarrow }\chi
_{q}^{2}$, where $q=rank(R)$ and $\hat{\Xi}\rightarrow _{p}\Xi =\Xi
_{1}^{-1}\Xi _{0}\Xi _{1}^{-1\prime }$.
\end{theorem}

The total number of regressions required to obtain the above break date
estimator $\hat{\lambda}$ is $O(T^{m})$. This can pose a considerable amount
of computational burden especially if the procedure is to be bootstrapped.
Hence, it is worthwhile to devise a test assuming diagonality in $\Sigma $
because the break dates are then estimated separately from each variable and
the total number of regressions required is only $O(T^{m_{1}}+\dots
+T^{m_{n}})$. While this modification lessens computational cost
significantly, the resulting break date estimators can be less efficient due
to the neglected correlation between variables in the system. Suppose the
break dates are estimated separately for each variable from (\ref{Model i}).
Thus, $\tilde{\lambda}=(\tilde{\lambda}_{1}^{\prime },\dots ,\tilde{\lambda}%
_{n}^{\prime })^{\prime }$ and%
\begin{equation}
\tilde{\lambda}_{i}=\arg \min\nolimits_{\lambda _{i}}SSR_{i}(\lambda _{i}),
\label{lambda_i_ols}
\end{equation}%
where $SSR_{i}(\lambda _{i})=y_{i}^{\prime }M_{\lambda _{i}}y_{i}$, with $%
M_{\lambda _{i}}=I_{T}-X(k_{i})(X(k_{i})^{\prime
}X(k_{i}))^{-1}X(k_{i})^{\prime }$. The limiting distribution of $\tilde{%
\lambda}$ is obtained as a special case of Theorem \ref{Theorem 2} by
letting $\Sigma =I$. The limiting covariance $\Xi $ can be simplified
somewhat. Define%
\begin{equation*}
\Xi _{s}=%
\begin{bmatrix}
\psi _{11}D_{\delta _{1}}^{-1}P_{11}D_{\delta _{1}}^{-1} & \dots & \psi
_{1n}D_{\delta _{1}}^{-1}P_{1n}D_{\delta _{n}}^{-1} \\ 
\vdots & \ddots & \vdots \\ 
\psi _{n1}D_{\delta _{n}}^{-1}P_{n1}D_{\delta _{1}}^{-1} & \dots & \psi
_{nn}D_{\delta _{n}}^{-1}P_{nn}D_{\delta _{n}}^{-1}%
\end{bmatrix}%
\text{,}
\end{equation*}%
where%
\begin{equation*}
P_{ij}=(\tint p_{i}(r)p_{i}^{\prime }(r)dr)^{-1}\tint p_{i}(r)p_{j}^{\prime
}(r)dr(\tint p_{j}(r)p_{j}^{\prime }(r)dr)^{-1},
\end{equation*}%
with $p_{i}(r)=g_{i}(r)-\tint g_{i}(r)f_{i}^{\prime }(r)dr(\tint
f_{i}(r)f_{i}^{\prime }(r)dr)^{-1}f_{i}(r)$ and $\psi _{ij}$ is the $%
(i,j)^{th}$ element of $\Psi $ defined in Assumption \ref{A3}.

\begin{corollary}
Suppose that Assumptions \ref{A1}$\sim $\ref{A3} hold. Let $\tilde{\lambda}=(%
\tilde{\lambda}_{1}^{\prime },\dots ,\tilde{\lambda}_{n}^{\prime })^{\prime
} $ where each $\tilde{\lambda}_{i}$ is obtained from (\ref{lambda_i_ols}).
Then, as $T\rightarrow \infty $: (i) $T^{3/2}(\tilde{\lambda}-\lambda ^{0})%
\overset{d}{\rightarrow }N(0,\Xi _{s})$; (ii) under $H_{0}$ in (\ref{H0}), $%
Wald=T^{3}(R\tilde{\lambda}-r)^{\prime }(R\hat{\Xi}_{s}R^{\prime })^{-1}(R%
\tilde{\lambda}-r)\overset{d}{\rightarrow }\chi _{q}^{2}$ where $q=rank(R)$
and $\hat{\Xi}_{s}\rightarrow _{p}\Xi _{s}$.\label{Corollary 1}
\end{corollary}

The result in (i) coincides with one result reported in Perron and Zhu
(2005) when there is only one equation with one break.

\section{Break Detection}

We consider a test for the null of $m=\sum_{i=1}^{n}m_{i}$ breaks in the
system against the alternative of $m+1$ breaks with the location of the
additional break unspecified. We again consider a quasi-likelihood
framework. Suppose that the model is now given by%
\begin{equation*}
y=X^{0}\theta +a_{i}(h)\gamma +u\text{,}
\end{equation*}%
where $a_{i}(h):=1_{i}\otimes b(h)$ for some $h\neq k_{i1},\dots ,k_{im_{i}}$
and $1_{i}$ is an $n\times 1$ vector with $1$ in the $i^{th}$ element and $0$
elsewhere. The null and alternative hypotheses are now:%
\begin{equation}
H_{0}:\gamma =0\ \ \ \ \mathrm{and}\ \ \ \ H_{1}:\gamma \not=0\text{.}
\label{H0 ab}
\end{equation}%
The maximum of the likelihood function under $H_{0}$ is again given by (\ref%
{loglike}) for a generic break date vector $k$. Similarly, the maximum of
the likelihood function under $H_{1}$ is:%
\begin{equation*}
l^{(i)}(k,h)=-(nT/2)(\log 2\pi +1)-(T/2)\log |\hat{\Sigma}_{(i)}(k,h)|\text{,%
}
\end{equation*}%
for a generic break date vector $k$ and an additional break date $h$. The
notation $l^{(i)}(k,h)$\ and $\hat{\Sigma}_{(i)}(k,h)$ is used to indicate
the fact that an additional break is inserted in the $i^{th}$ equation. This
additional break can only occur in one of the equations and we assume for
simplicity that the relevant equation is known. This assumption is relaxed
below. Just like we have defined $\lambda =k/T$ and used $l(k)$ and $%
l(\lambda )$ interchangeably, we define $\nu =h/T$ and will use $%
l^{(i)}(k,h) $ and $l^{(i)}(\lambda ,\nu )$ interchangeably. The test
statistic we consider is given by%
\begin{equation*}
LR=-2[l(\hat{\lambda})-\sup\nolimits_{\nu \in C_{T}^{(i)}}l^{(i)}(\hat{%
\lambda},\nu )],
\end{equation*}%
where $\hat{\lambda}=\arg \sup_{\lambda }l(\lambda )$ and%
\begin{equation*}
C_{T}^{(i)}=\left\{ 
\begin{array}{c}
\varepsilon _{ps}\leq \nu \leq 1-\varepsilon _{ps}\text{ and }|\nu -\hat{%
\lambda}_{ij}|\geq \varepsilon _{ps}\text{, }j=1,\dots ,m_{i} \\ 
\text{ for some }\varepsilon _{ps}\text{, }0<\varepsilon _{ps}<\min \{\hat{%
\lambda}_{i1},\hat{\lambda}_{im_{i}}\}\text{.}%
\end{array}%
\right\} .
\end{equation*}%
As is well known in the structural break literature, $\gamma $ is not
identified under the null hypothesis and $\nu $ must be restricted to ensure
a non-divergent limiting distribution of the test statistic. The set $%
C_{T}^{(i)}$ imposes the relevant restrictions.\footnote{%
Instead of using $\hat{\lambda}$ under both $H_{0}$ and $H_{1}$, it is
possible to jointly minimize $(\lambda ,\nu )$ under $H_{1}$. However, this
requires a more complex asymptotic analysis and we will focus on the simpler
case.} When the equation with an additional break is not specified a priori,
the $LR$ statistic can be extended to be%
\begin{equation*}
\sup LR=-2[l(\hat{\lambda})-\max\nolimits_{1\leq i\leq n}\sup\nolimits_{\nu
\in C_{T}^{(i)}}l^{(i)}(\hat{\lambda},\nu )]\text{.}
\end{equation*}%
In order to express the limiting distribution, we need to define additional
terms. Let $f_{i}(r)$ and $g_{i}(r)$ be as defined before. Let $b(r,\nu
)=(r-\nu )^{+}$, $FB_{i}(\nu )=\int f_{i}(r)b(r,\nu )dr$, $GB_{i}(\nu )=\int
g_{i}(r)b(r,\nu )dr$ and $BB(\nu )=\int b^{2}(r,\nu )dr$ where all integrals
are taken from 0 to 1. Recall that $s_{ij}$ is the $(i,j)^{th}$ element of $%
\Sigma ^{-1}$ and $\kappa _{ij}$ is that of $\Sigma ^{-1}\Psi \Sigma ^{-1}$,
and let $Q_{BB}^{(i)}(\nu )=s_{ii}BB(\nu )$, $\Gamma _{BB}^{(i)}(\nu
)=\kappa _{ii}BB(\nu )$, $Q_{FB}^{(i)}(\nu )=(s_{1i}FB_{1}(\nu )^{\prime
},...,s_{ni}FB_{n}(\nu )^{^{\prime }})^{\prime }$, $\Gamma _{FB}^{(i)}(\nu
)=(\kappa _{1i}FB_{1}(\nu )^{^{\prime }},...,\kappa _{ni}FB_{n}(\nu
)^{^{\prime }})^{\prime }$, as well as $Q_{GB}^{(i)}(\nu )=(s_{1i}GB_{1}(\nu
)^{^{\prime }},...,s_{ni}GB_{n}(\nu )^{^{\prime }})^{\prime }$ and $\Gamma
_{GB}^{(i)}(\nu )=(\kappa _{1i}GB_{1}(\nu )^{^{\prime }},...,\kappa
_{ni}GB_{n}(\nu )^{^{\prime }})^{\prime }$. Now, using these functions, we
define%
\begin{eqnarray*}
\xi _{0}^{(i)}(\nu ) &=&\Gamma _{BB}^{(i)}(\nu )-Q_{BF}^{(i)}(\nu
)Q_{FF}^{-1}\Gamma _{FB}^{(i)}(\nu )-\Gamma _{BF}^{(i)}(\nu
)Q_{FF}^{-1}Q_{FB}^{(i)}(\nu )+Q_{BF}^{(i)}(\nu )Q_{FF}^{-1}\Gamma
_{FF}Q_{FF}^{-1}Q_{FB}^{(i)}(\nu ) \\
\xi _{1}^{(i)}(\nu ) &=&Q_{BB}^{(i)}(\nu )-Q_{BF}^{(i)}(\nu
)Q_{FF}^{-1}Q_{FB}^{(i)}(\nu ) \\
\varsigma _{0}^{(i)}(\nu ) &=&D_{\delta }[\Gamma _{GB}^{(i)}(\nu
)-Q_{GF}Q_{FF}^{-1}\Gamma _{FB}^{(i)}(\nu )-\Gamma
_{GF}Q_{FF}^{-1}Q_{FB}^{(i)}(\nu )+Q_{GF}Q_{FF}^{-1}\Gamma
_{FF}Q_{FF}^{-1}Q_{FB}^{(i)}(\nu )] \\
\varsigma _{1}^{(i)}(\nu ) &=&D_{\delta }[Q_{GB}^{(i)}(\nu
)-Q_{GF}Q_{FF}^{-1}Q_{FB}^{(i)}(\nu )]\text{,}
\end{eqnarray*}%
where $D_{\delta }$, $Q_{GF}$, $Q_{FF}$, $\Gamma _{GF}$ and $\Gamma _{FF}$
are as previously defined. Finally, let $\eta _{(i)}(\nu )$ be a mean zero
Gaussian process defined over the unit interval with%
\begin{equation*}
Var(\eta _{(i)}(\nu ))=\xi _{0}^{(i)}(\nu )-\varsigma _{1}^{(i)}(\nu
)^{\prime }\Xi _{1}^{-1}\varsigma _{0}^{(i)}(\nu )-\varsigma _{0}^{(i)}(\nu
)^{\prime }\Xi _{1}^{-1}\varsigma _{1}^{(i)}(\nu )+\varsigma _{1}^{(i)}(\nu
)^{\prime }\Xi _{1}^{-1}\Xi _{0}\Xi _{1}^{-1}\varsigma _{1}^{(i)}(\nu )\text{%
.}
\end{equation*}%
Also, denote the limit counterpart of the set $C_{T}^{(i)}$ by $C^{(i)}$,
replacing $\hat{\lambda}_{ij}$ by $\lambda _{ij}^{0}$. The limit
distribution of the tests are stated in the following theorem.

\begin{theorem}
\label{Theorem 3} Suppose that Assumptions \ref{A1}$\sim $\ref{A3} hold.
Then, as $T\rightarrow \infty $, we have under $H_{0}$ in (\ref{H0 ab}): (i) 
$LR\overset{d}{\rightarrow }\sup_{\nu \in C^{(i)}}[\eta _{(i)}^{2}(\nu )/\xi
_{1}^{(i)}(\nu )]$; (ii) $\sup LR\overset{d}{\rightarrow }\sup_{1\leq i\leq
n}\sup_{\nu \in C^{(i)}}[\eta _{(i)}^{2}(\nu )/\xi _{1}^{(i)}(\nu )]$.
\end{theorem}

The limiting distributions depend on various nuisance parameters. First, all
of the true break fractions $\lambda _{ij}^{0}$ for $j=1,\dots ,m_{i}$ and $%
i=1,\dots ,n$ matter via terms such as $Q_{FB}^{(i)}(\nu )$, $%
Q_{GB}^{(i)}(\nu )$, $\Gamma _{FB}^{(i)}(\nu )$, $\Gamma _{GB}^{(i)}(\nu )$, 
$Q_{FF}$, $Q_{GF}$, $\Gamma _{FF}$ and $\Gamma _{GF}$. Second, both the
short-run and long-run variances matter via the various $Q$ and $\Gamma $
matrices. Third, the trimming parameter $\varepsilon _{ps}$ in the set $%
C^{(i)}$ matters. Given the complexity of the limiting distributions, we
neither seek a way to eliminate nuisance parameters nor attempt to tabulate
critical values. Instead, we resort to using a bootstrap method to generate
p-values. The various terms are evaluated in a manner similar to that of the
tests for common breaks described in the next section. Note that if the
short-run and long-run variances coincide and the true break fractions $%
\lambda _{ij}^{0}$ are used instead of their estimates, $Var(\eta _{(i)}(\nu
))$ reduces to be $\xi _{1}^{(i)}(\nu )$.

\section{Monte Carlo Simulations}

In this section we provide Monte Carlo simulation results to assess the
adequacy of the asymptotic distributions derived in Sections 3 and 4. All
results are based on 1,000 replications. We focus on the test for common
breaks as the tests for an additional change in slope are too
computationally demanding so that a reasonable simulation experiment is
prohibitive. The data generating process (DGP) is specified by:%
\begin{eqnarray*}
y_{1t} &=&\mu _{1}+\beta _{1}t+\delta _{11}b_{t}(k_{11}^{0})+u_{1t} \\
y_{2t} &=&\mu _{2}+\beta _{2}t+\delta _{21}b_{t}(k_{21}^{0})+u_{2t}\text{.}
\end{eqnarray*}%
The results are exactly invariant to the values of $\mu _{i}$ and $\beta
_{i} $. For the slope change parameters $\delta _{11}$ and $\delta _{21}$,
the cases of 0.5, 1.0 and 1.5 are considered. The error terms are such that%
\begin{equation*}
\begin{pmatrix}
u_{1t} \\ 
u_{2t}%
\end{pmatrix}%
=L%
\begin{pmatrix}
e_{1t} \\ 
e_{2t}%
\end{pmatrix}%
\text{ \ with \ }LL^{\prime }=%
\begin{pmatrix}
1 & \rho \\ 
\rho & 1%
\end{pmatrix}%
\text{,}
\end{equation*}%
and $(e_{1t},e_{2t})^{\prime }=\alpha (e_{1t-1},e_{2t-1})^{\prime
}+(\varepsilon _{1t},\varepsilon _{2t})^{\prime }$, with $(\varepsilon
_{1t},\varepsilon _{2t})^{\prime }\sim i.i.d.$ $N\left( 0,(1-\alpha
)^{2}I_{2}\right) $. The parameter $\rho $ stands for the correlation across
equations and $\alpha $ stands for the autoregressive parameter in each of
the error terms. We consider (-0.5, 0, 0.5) for the values of $\rho $ and
(0, 0.3, 0.7) for the values of $\alpha $. What influences the precision of
the break date estimate is the ratio between the break magnitude and the
long-run variance. In our simulations, the long-run variance is always set
to be unity. The sample size is set to $T=100$ and the true break dates are
at mid-sample, $k_{11}^{0}=k_{21}^{0}=50$. For each generated data, we test
four null hypotheses: (1) $R=I_{2}$ and $r=(0.5,0.5)^{\prime }$, (2) $%
R=I_{2} $ and $r=\left( 0.525,0.475\right) ^{\prime }$, (3) $R=[1,-1]$ and $%
r=0$ and (4) $R=[1,-1]$ and $r=0.05$. Hence the null rejection probabilities
for (1) and (3) pertain to the finite sample sizes (with a nominal size 5\%)
while those for (2) and (4) pertain to powers. In testing these hypotheses,
we consider the LR and Wald tests based on the SUR system (Theorems \ref%
{Theorem 1} and \ref{Theorem 2}) as well as the Wald test based on
estimating the break dates equation by equation using OLS regressions
(Corollary \ref{Corollary 1}). We refer to the two Wald tests as the
GLS-Wald and OLS-Wald, respectively. For both versions, the long-run
variance is estimated using a quadratic spectral window, where the bandwidth
parameter is selected using Andrews' (1991) data dependent method with an
AR(1) approximation.

We also simulate the bootstrap version of the aforementioned tests. To
obtain bootstrap p-values, we first estimate break dates imposing the null
hypothesis and remove the estimated trend. Then, we fit a VAR(1) on the
resulting residuals. Following Kilian (1998), we compute the bias-corrected
estimates of the VAR coefficients and obtain the corresponding residuals of
the VAR. From them, we construct a pseudo VAR process and add to it the
estimated trend functions to obtain a bootstrap sample. Simulating bootstrap
tests requires a lot of computing time especially for those based on the
full system. We make use of the Warp-Speed method suggested by Giacomini,
Politis and White (2013). Also, we use the feasible GLS estimate instead of
the MLE in the construction of the LR and GLS-Wald tests.

In Table 1, the results for the LR test are reported. Regarding the finite
sample size obtained using the asymptotic critical values, two features
emerge. First, the size is near or below the nominal level when $\alpha =0$,
but it climbs up quickly as $\alpha $ increases. In the worst case, it can
go up to 50\%. This result is well expected since the asymptotic validity of
the LR test holds only when $\alpha =0$. Second, the size inflation is more
evident when $\delta _{i}$ is small. In fact, when $\delta _{i}$ is very
large, the estimated break dates coincide with the true ones making the test
statistic literally zero with a large probability. However, this type of
conservativeness is of little concern because the power function does not
appear to be decreasing as $\delta _{i}$ gets large. On the other hand, the
finite sample size obtained by the bootstrap procedure stays near the
nominal 5\% level across all simulation designs, offering significant
improvements over the asymptotic test. The bootstrapped test gives slightly
smaller power than the asymptotic test, but the difference is marginal
especially in view of the large improvement in size.

Table 2 reports the results for the GLS-Wald test. Overall, the GLS-Wald
test using the asymptotic critical values are more liberal than the
corresponding LR test, which is a well known characteristic of Wald tests.
Despite being liberal, the GLS-Wald is less sensitive to the value of $%
\alpha $, since we are using a robust covariance matrix. The GLS-Wald test
also gets conservative as $\delta _{i}$ gets large for the same reason as
the LR test. The bootstrapped GLS-Wald test controls the size as well as the
bootstrapped LR test. Table 3 shows the results for the OLS-Wald test. It
performs similarly to the GLS-Wald test for both the asymptotic and
bootstrap versions. Lastly, the power is close to one in almost all cases.
The hypotheses that are rejected misspecifies the break dates by only 5\% of
the sample size, yet our tests are powerful enough to detect them. To sum
up, all three tests, the LR, GLS-Wald, and OLS-Wald suffer from size
distortions if the breaks are not large enough. However, the bootstrap
procedures significantly help control the size without losing much power.

\section{Applications}

We investigate the commonality of the break dates across the temperature and
anthropogenic total radiative forcing series as well as test whether the
recent ``hiatus''\ is significant. We use two sets of global, northern and
southern hemispheric temperature series, each of which will be denoted as G,
N and S, respectively. The first set comes from the Climate Research Unit's
HadCRUT4 (Morice et al., 2012) and the second from the NASA database
(GISTEMP Team, 2015; Hansen et al., 2010). For global temperatures, we also
use the data from Berkeley Earth (Rohde et al., 2013) and the dataset in
Karl et al. (2015). We first discuss the data used, then the results for the
common breaks tests and finally the results for the test on a change in
slope in temperatures related to the so-called ``hiatus'' period.

\subsection{The data}

The annual temperature data used are from the HadCRUT4 (1850-2014)
(http://www.metoffice. gov.uk /hadobs/hadcrut4/data/current/download.html)
and the GISS-NASA (1880-2014) datasets (http://data.giss.nasa.gov/gistemp/).
The Atlantic Multidecadal Oscillation (AMO) and the North Atlantic
Oscillation (NAO) series (1856-2014) are from NOAA; http://www.esrl
.noaa.gov/psd/data/timeseries/AMO/ and http://www.esrl.noaa.gov/psd/
gcos\_wgsp/ Timeseries/NAO/). As stated above, for global temperatures, we
also use the data from Berkeley Earth (Rohde et al., 2013) and the dataset
in Karl et al. (2015). We also use series from databases related to climate
model simulations by the Goddard Institute for Space Studies (GISS-NASA).
The radiative forcing data obtained from GISS-NASA
(https://data.giss.nasa.gov/ modelforce/; Hansen et al., 2011) for the
period 1880-2010 include the following (in W/m2): well-mixed greenhouse
gases, WMGHG, (carbon dioxide, methane, nitrous oxide and
chlorofluorocarbons); ozone; stratospheric water vapor; solar irradiance;
land use change; snow albedo; stratospheric aerosols; black carbon;
reflective tropospheric aerosols; and the indirect effect of aerosols. The
aggregated radiative forcing series are constructed as follows: WMGHG is the
radiative forcing of the well-mixed greenhouse gases and has a largely
anthropogenic origin; Total Radiative Forcing (TRF) is WMGHG plus the
radiative forcing of ozone, stratospheric water vapor, land use change; snow
albedo, black carbon, reflective tropospheric aerosols, the indirect effect
of aerosols and solar irradiance.

\subsection{Common breaks tests}

The temperature series are affected by various modes of natural variability
such as the Atlantic Multidecadal Oscillation (AMO) and the North Atlantic
Oscillation (NAO), which are characterized by low frequency movements (Kerr,
2000; Hurrell, 1995).\ Since trends and breaks are low frequency features,
it is important to purge them from the temperature series allowing more
precise estimates of the break dates.\footnote{%
We also considered filtering the effect of the Pacific Decadal Oscilliation
(PDO) index (https://www.ncdc.noaa.gov/teleconnections/pdo/) and the
Southern Oscilliation Index (SOI) (http://www.cru.uea.ac.uk/cru/data/soi/);
see Wolter and Timlin (1998) and Mantua and Hare (2002). The results were
qualitatively similar showing robustness. Hence, we \ do not report them.}
Other high frequency fluctuations in temperature series do not affect the
precision of the estimates of the break dates and the magnitudes of the
changes in slope. Accordingly, we filter out the effect of these modes of
variability by regressing each temperature series on these modes and a
constant. Since the effect of natural variability might have occurred with a
time lag, we choose an appropriate lag using the Bayesian Information
Criterion (BIC); Schwarz (1978). The candidate regressors for the filtering
are the current value and lags (up to order $kmax-1$) of AMO and NAO. We
first work with G from HadCRUT4. We start with $kmax=2$, so that the
candidates are the current value and the first lag only. BIC chooses the
current value of AMO and the first lag of NAO. Since the maximum lag allowed
is selected, we increase $kmax$ to 4. Then, BIC chooses the current value of
AMO and the second lag of NAO (the first is not included given that we
search over models not necessarily including all lags up to some chosen
order). When applying the BIC, the number of observations used is limited by 
$kmax$ (e.g., Perron and Ng, 2005). Having decided on the current value of
AMO and the second lag of NAO, we apply the filtering to all available
observations, not limited by $kmax$. We could repeat the same procedure to
each series. However, it does not make much sense to have the same mode
affect each temperature series with a different lag. Hence, we filter all
series with the current value of AMO and the second lag of NAO.\footnote{%
If we choose the filtering regressors for each series, G and SH from NASA
and G from Berkeley require the fourth lag of NAO instead of the second. But
the break date estimates change by two years at most.} The filtered
temperature series are denoted as \~{G}, \~{N} and \~{S}. Figure 1 presents
graphs of the original and filtered series.

For the radiative forcing variables, we use WMGHG and TRF. Figure 2 displays
the time plots of WMGHG and TRF as well as the AMO and NAO. The data set
used covers the period 1880-2014 and the filtering is done with the full
sample. We use the data for 1900-2014 for the purpose of the common break
date tests.\footnote{%
For the radiative forcing variables, the data is available only up to 2011.
We use forecast values for 2012-2014 based on an autoregressive model with a
broken linear trend. In the case of TRF the forecast is produced using an
AR(2) and a broken linear trend with 1960 and 1991 as the break dates. For
WMGHG, we use an AR(8) and and a broken linear trend with break dates at
1960 and 1994.} We use two sample periods: 1900-1992 and 1963-2014. The
reason we split the sample is to reduce the number of breaks in the system,
thereby lessening the computational burden. Our choice of sample periods is
made with the intention to minimize any unwanted effect from a second break
that might exist in any of the variables in the system. The break dates in
WMGHG are estimated most precisely since it has a very small noise component
(see Figure 2) and they are 1963 and 1992. First we consider bivariate
systems with one temperature series and one radiative forcing series. The
null hypothesis of a common break date is tested using the LR test from
Theorem \ref{Theorem 1}. Since this test is not asymptotically pivotal, we
supply bootstrap p-values as well. The bootstrap is carried out in the same
way as done in the Monte Carlo simulations reported earlier. We also report
the GLS-Wald test from Theorem \ref{Theorem 2}. Although the GLS-Wald test
is asymptotically pivotal, we still supply the bootstrap p-values because of
the size distortions observed in our Monte Carlo study. In addition to the 4
global, 2 northern hemispheric and 2 southern hemispheric temperature
series, we also consider the average of the global, northern and southern
hemispheric temperature series across different datasets. Hence, there are
11 temperature series and 11 filtered temperature series, each of which is
paired with either WMGHG or TRF, yielding a total of 44 bivariate systems.

Table 4 (a) reports the results for the sample period of 1900-1992 when the
original temperature series are considered; the asymptotic p-values of the
Wald and LR test are very small in most cases and the null of a common break
date is rejected with only a few exceptions. However, the bootstrap p-values
are well over the conventional thresholds 0.05 or 0.01 in all cases. Given
the previous simulation results, we view the bootstrap p-values as more
reliable than the asymptotic ones. Note that the break date for the
radiative forcing variable is almost unchanged at either 1963 or 1966,
regardless of the temperature series in the bivariate system. Also, the
estimated common break date, i.e., the estimate imposing the null of common
breaks, always coincide with the break date for the radiative forcing
variable. On the other hand, the break date estimates for the temperature
series are quite spread out. Hence, the break dates in the radiative forcing
variables are being estimated more precisely than those in the temperature
series. The large p-values produced by the bootstrapping is a consequence of
the high level of noise in the temperature series relative to the magnitude
of the break. Accordingly, we fail to reject the null of a common break date
despite the large discrepancy in the estimated break dates. Table 4 (a) also
shows the results obtained with the filtered temperature series. The picture
changes quite dramatically. The break dates for the radiative forcing
variables are still at 1963 or 1966, but those for the temperature series
are now also around 1960. Especially, when TRF enters the bivariate system,
the estimate of the break date is 1960, 1965 or 1966, the exception being
the southern hemispheric temperature from HadCRUT4 and the average of the
two southern hemispheric temperature series, for which the estimate is still
1956. As a result, both the asymptotic and bootstrap p-values are much
larger than those for the original temperature series.

Tables 4 (b) presents the results for the second sample 1963-2014 when the
original temperature series are used and the estimates of the break dates
for the temperature series are again quite spread out, while those for the
radiative forcing variables are consistently around 1990-1992. Just like in
Table 4 (a), the asymptotic p-values suggest strong rejection of the null
hypothesis, but the bootstrap p-values turns out much larger reflecting
again the high level of correlated noise in the original series. Note that
the estimates of the break dates for the filtered temperature series are
1990-1992 with only a few exceptions. Also, the p-values are larger than the
corresponding ones for the original temperature series.

We repeat the analysis for the bivariate systems with the OLS-Wald test as a
robustness check and report the results in Table 5. The results are very
similar when the filtered series are considered. The break date estimates
obtained equation by equation coincide with those obtained from the system
or differ only by a few years at most. When the original series are used,
the break date estimates for some series largely deviate from those computed
by the system method. However, the fact that the bootstrap p-values are much
larger than usual thresholds still applies. Thus, our conclusions remain the
same.

An advantage of the OLS-Wald test is that it can handle a larger system
without extra computational burden. The break date estimates will not change
if we consider a larger system, because they are computed equation by
equation. However, the long-run variance estimate and the bootstrap p-values
can change as we consider a bigger system. We consider three five-variables
systems. Each system has a global, northern and southern hemispheric
temperature and the two radiative forcing variables. The first one uses the
temperature series from HadCRUT4, the second one from NASA and the third one
has average global, northern and southern hemispheric temperature series.
The results are reported in Table 6. They show results broadly similar to
those obtained bivariate systems.

To provide a final robustness check, we consider testing jointly whether the
two breaks near 1960 and for the ``hiatus'' are common. We use the OLS-Wald
test, which estimates the two breaks jointly for each equation, and the
period 1926-2014 to focus on the time span whereby anthropogenic factors are
most likely to be the main driver affecting changes in temperatures. The
results are presented in Table 7. The results obtained are broadly similar
and lead to the same conclusions. When using the filtered series and the
bootstrap p-values, the test fails to reject the null hypothesis of two
common breaks. The first is located in the early 60's (the onset of
sustained global warming) and the second in the early 90's (the ``hiatus'')
consistent with methods dealing with one break at a time presented earlier.

\subsection{Testing for the ``hiatus''}

The literature on the existence and drivers of the current slowdown in the
rate of warming has expanded quickly (Tollefson, 2014, 2016). In this
section, we focus on testing if the ``hiatus'' can be explained by natural
variability or if it is a feature of the underlying warming trend (Estrada,
Perron and Martinez, 2013). A small part of the literature investigating the
slowdown in the warming in global temperatures are based on formal
structural change/change-point tests (Cahill et al., 2015; Pretis et al.,
2015), while others are simply based on testing for trends during
arbitrarily selected periods of time (Cahill et al., 2015; Foster and
Rahmstorf, 2011; Lewandowsky et al., 2015). This type of approach, as well
as most of physical-based studies, have typically failed to find convincing
evidence for the existence of a change in the slope of global temperatures
during the last decades. According to one of the commonly accepted
hypothesis, the apparent ``hiatus'' is produced by the effects of
low-frequency natural variability that result from coupled ocean-atmosphere
processes and heat exchange between the ocean and the atmosphere. It has
been proposed that effects of natural variability modes such as AMO, NAO and
PDO were able to mask the warming trend since the 1990s, creating the
illusion of a slowdown in the underlying warming trend (e.g., Guan et al .,
2015; Steinman et al., 2015; Li, Sun and Jin, 2013; Trenberth and Fasullo,
2013). However, Estrada and Perron (2016) argue that low-frequency
oscillations do distort the underlying warming trend but cannot account for
the current slowdown. Instead, the main effect of these oscillations has
been to make it more difficult to detect the drop in the rate of warming, a
real feature of the warming trend imparted by the slowdown in the radiative
forcing from well-mixed greenhouse gases.

Establishing whether the ``hiatus'' period is statistically significant is
important because it would counter the widely held view that it is the
product of natural internal variability (Kosaka and Xie, 2013; Trenberth and
Fasullo, 2013; Meehl et al., 2011; Balmaseda, Trenberth and K\"{a}ll\'{e}n,
2013). Using standard tests (e.g., Perron and Yabu, 2009), the results are
mixed across various series and sometimes borderline. Our aim is to provide
tests with enhanced power by casting the testing problem in a bivariate
framework involving temperatures and radiative forcing. We consider
bivariate systems with one temperature series and one forcing variable. We
use the LR test for the presence of a break in temperature series given the
presence of a structural break in radiative forcing series. We first
estimate a break date in the radiative forcing series under the likelihood
framework for a bivariate system of a pair of radiative forcing and
temperature series. Then, given the estimated break date in the radiative
forcing series, we apply the LR test for the null hypothesis of no break in
temperatures versus one break. To obtain $p$-values, we apply the bootstrap
procedure described in Section 5 with 1,000 samples. The results are
presented in Table 8. First, as expected from previous results, the LR test
is more likely to reject the null hypothesis of no break when using filtered
temperature series. Hence, we report results only for that case. For the
filtered series from 1900 to 1992, we reject the null at less than 5\%
significance level in all pairs of radiative forcing and temperature series.
Hence, there is clear evidence of a break in temperatures that is near 1960
(varying between 1954 and 1966 depending on the series and the forcing).
This concurs with the common break results obtained in the previous section.
When using the sample 1963-2014 we reject the null in seven and eight cases
out of eleven filtered series at the 10\% significance level with well-mixed
green-house gases and total radiating forcing, respectively. Since the
sample size is small ($T=52)$, we might expect that the LR test may have
little power, but our result suggests strong evidence for the presence of a
break in temperature series. Note that the evidence for a break is stronger
when using bivariate systems involving TRF. This is due to the fact that TRF
exhibits a larger decrease in slope compared to WMGHG. The errors are also
more strongly correlated. For instance, solar irradiance is a part of both
TRF and temperatures and an important source of variations. When considering
systems with TRF, the only pairs that do not allow a rejection at the 10\%
level are those associated with Northern hemisphere temperatures. For these,
the p-values range from .14 to .19, while for all other pairs they are below
10\%. Given that the small sample size is small, the overall evidence
strongly indicates a break in temperatures in the early 90s consistent with
the presence of the much-debated ``hiatus''.

\section{Conclusion}

We consider a multivariate system with $n$ equations with the dependent
variables modeled as joint-segmented trends with multiple changes in slope.
The errors can be serially correlated and correlated across equations. We
consider testing for general linear restrictions on the break dates,
including testing for common breaks across equations. The test used is a
(quasi-) likelihood ratio test assuming serially uncorrelated errors. Under
the stated conditions, the LR\ test has a pivotal chi-square distribution,
though it is non-pivotal in general cases of interest. Hence, we also
consider a corrected Wald test having a pivotal limit distribution, which
can be constructed using break dates estimated one equation at a time or via
the complete system. The limit distribution of the test is standard
chi-square. However, simulations show that the two Wald tests suffer from
potentially severe liberal size distortions for moderate to strong serial
correlation. Hence, for all three tests, we suggest a bootstrap procedure to
obtain the relevant critical values. These bootstrap tests are shown to have
correct size in all cases considered. Our empirical results show that, once
we filter the temperature data for the effect of the Atlantic Mutidecadal
Oscillation (AMO) and the North Atlantic Oscillation (NAO), the breaks in
the slope of radiative forcing and temperatures are common, both for the
large increase in the 60s and the recent ``hiatus''.

We also consider a test for the presence of an additional break in some
series. The theoretical framework is general and allows multiple breaks in a
general multivariate system. The test considered is a quasi-likelihood ratio
test. The limit distribution is shown to be non-pivotal and depends in a
complex way on a number of nuisance parameters. Hence, we use a bootstrap
procedure to obtain relevant critical values. In our applications, the null
and alternative hypotheses specify a break in the slope of the radiative
forcing series, while under the null hypothesis there is no break in the
temperature series but there is one under the alternative hypothesis. Our
results indicate that indeed the ``hiatus''\ represents a significant
slowdown in the rate of increase in temperatures, especially when
considering global or southern hemisphere series, for which our test points
to a rejection of the null of no change for all data sources considered. The
statistical results are of independent interest and applicable more
generally beyond the climate change applications considered.

\baselineskip=18.0pt\centerline{\bf Appendix} \setcounter{section}{0} %
\setcounter{equation}{0} \setcounter{lemma}{0}\setcounter{page}{1} %
\renewcommand{\thepage}{A-\arabic{page}} \renewcommand{\theequation}{A.%
\arabic{equation}}\renewcommand{\thelemma}{A.\arabic{lemma}} \baselineskip%
=14pt\bigskip

\textbf{Proof of Theorems \ref{Theorem 1} and \ref{Theorem 2}:} Let $\hat{%
\lambda}_{r}=\arg \min_{\lambda \text{ s.t. }R\lambda =r}|\hat{\Sigma}%
(\lambda )|$ and $\hat{\lambda}=\min_{\lambda }|\hat{\Sigma}(\lambda )|$.
Given the consistency of $\hat{\Sigma}(\hat{\lambda}_{r})$, $\hat{\Sigma}(%
\hat{\lambda})$ and $\hat{\Sigma}(\lambda ^{0})$ under the null hypothesis,
we can apply a Taylor series expansion to $LR$ such that%
\begin{eqnarray}
LR &=&T~tr[\hat{\Sigma}^{-1}(\hat{\lambda})(\hat{\Sigma}(\hat{\lambda}_{r})-%
\hat{\Sigma}(\hat{\lambda}))]+o_{p}(1)  \notag \\
&=&T~tr[\hat{\Sigma}^{-1}(\hat{\lambda})(\hat{\Sigma}(\hat{\lambda}_{r})-%
\hat{\Sigma}(\lambda ^{0}))]-T~tr[\hat{\Sigma}^{-1}(\hat{\lambda})(\hat{%
\Sigma}(\hat{\lambda})-\hat{\Sigma}(\lambda ^{0}))]+o_{p}(1).
\label{LR decomposition}
\end{eqnarray}%
Consider the first term in the above decomposition. We may write%
\begin{gather*}
T~tr[\hat{\Sigma}^{-1}(\hat{\lambda})\left( \hat{\Sigma}(\hat{\lambda}_{r})-%
\hat{\Sigma}(\lambda ^{0})\right) ]=tr[\hat{\Sigma}^{-1}(\hat{\lambda})(%
\hat{U}_{\hat{\lambda}_{r}}^{\prime }\hat{U}_{\hat{\lambda}_{r}}-\hat{U}%
_{\lambda ^{0}}^{\prime }\hat{U}_{\lambda ^{0}})] \\
=tr[\Sigma ^{-1}(\tilde{U}_{\hat{\lambda}_{r}}^{\prime }\tilde{U}_{\hat{%
\lambda}_{r}}-\tilde{U}_{\lambda ^{0}}^{\prime }\tilde{U}_{\lambda
^{0}})]+o_{p}(1)=\min_{\lambda \text{ s.t. }R\lambda =r}tr[\Sigma ^{-1}(%
\tilde{U}_{\lambda }^{\prime }\tilde{U}_{\lambda }-\tilde{U}_{\lambda
^{0}}^{\prime }\tilde{U}_{\lambda ^{0}})]+o_{p}(1),
\end{gather*}%
where $\tilde{U}_{\lambda }$ is the matrix of estimated residuals from the
infeasible GLS estimator, instead of the ML estimator, for $\lambda $. For
notational simplicity, write $\Omega =\Sigma \otimes I_{T}$ and $M(\lambda
)=I-\Omega ^{-1/2}X(X^{\prime }\Omega ^{-1}X)^{-1}X^{\prime }\Omega ^{-1/2}$%
. Then the test statistic admits the following decomposition:%
\begin{gather}
tr[\Sigma ^{-1}(\tilde{U}_{\lambda }^{\prime }\tilde{U}_{\lambda }-\tilde{U}%
_{\lambda ^{0}}^{\prime }\tilde{U}_{\lambda ^{0}})]=vec(\tilde{U}_{\lambda
})^{\prime }\Omega ^{-1}vec(\tilde{U}_{\lambda })-vec(\tilde{U}_{\lambda
^{0}})^{\prime }\Omega ^{-1}vec(\tilde{U}_{\lambda ^{0}})  \notag \\
=\theta ^{\prime }(X^{0}-X)^{\prime }\Omega ^{-1/2\prime }M(\lambda )\Omega
^{-1/2}(X^{0}-X)\theta +2\theta ^{\prime }(X^{0}-X)^{\prime }\Omega
^{-1/2\prime }M(\lambda )\Omega ^{-1/2}u  \notag \\
+u^{\prime }\Omega ^{-1/2\prime }\left[ M(\lambda )-M(\lambda ^{0})\right]
\Omega ^{-1/2}u  \notag \\
\equiv (XX)+2(XU)+(UU)\text{.}  \label{XX+2XU+UU}
\end{gather}%
Note that%
\begin{eqnarray*}
\left[ X(k_{i}^{0})-X(k_{i})\right] \theta _{i}
&=&[0,0,b(k_{i1}^{0})-b(k_{i1}),\dots
,b(k_{im_{i}}^{0})-b(k_{im_{i}})]\theta _{i} \\
&=&[0,0,(k_{i1}-k_{i1}^{0})\tilde{\iota}(k_{i1}),\dots
,(k_{im_{i}}-k_{im_{i}}^{0})\tilde{\iota}(k_{im_{i}})]\theta _{i} \\
&=&[(k_{i1}-k_{i1}^{0})\tilde{\iota}(k_{i1}),\dots
,(k_{im_{i}}-k_{im_{i}}^{0})\tilde{\iota}(k_{im_{i}})]\delta _{i}\equiv 
\tilde{\iota}(k_{i})diag(\delta _{i})(k_{i}-k_{i}^{0}),
\end{eqnarray*}%
where $\delta _{i}=(\delta _{i1},\dots ,\delta _{im_{i}})^{\prime }$, $%
\tilde{\iota}(k_{ij})=(k_{ij}-k_{ij}^{0})^{-1}(b(k_{ij}^{0})-b(k_{ij}))$, $%
k_{i}=(k_{i1},\dots ,k_{im_{i}})^{\prime }$ and $\tilde{\iota}(k_{i})=[%
\tilde{\iota}(k_{i1}),\dots ,\tilde{\iota}(k_{im_{i}})]$. Now,%
\begin{eqnarray*}
\left[ X^{0}-X\right] \theta  &=&%
\begin{pmatrix}
\left[ X(k_{1}^{0})-X(k_{1})\right] \theta _{1} &  & 0 \\ 
& \ddots  &  \\ 
0 &  & \left[ X(k_{n}^{0})-X(k_{n})\right] \theta _{n}%
\end{pmatrix}
\\
&=&%
\begin{pmatrix}
\tilde{\iota}(k_{1}) &  & 0 \\ 
& \ddots  &  \\ 
0 &  & \tilde{\iota}(k_{n})%
\end{pmatrix}%
\begin{pmatrix}
D_{\delta _{1}} &  & 0 \\ 
& \ddots  &  \\ 
0 &  & D_{\delta _{n}}%
\end{pmatrix}%
\begin{pmatrix}
k_{1}-k_{1}^{0} \\ 
\vdots  \\ 
k_{n}-k_{n}^{0}%
\end{pmatrix}%
\equiv \Psi (\delta )(k-k^{0}),
\end{eqnarray*}%
with $D_{\delta _{i}}=diag(\delta _{i})$. Thus%
\begin{eqnarray*}
\left| (XX)\right|  &\leq &\left\| \Omega ^{-1/2}(X^{0}-X)\theta \right\|
^{2}\leq \left\| \Omega ^{-1/2}\right\| ^{2}\left\| (X^{0}-X)\theta \right\|
^{2} \\
&=&\left\| \Omega ^{-1/2}\right\| ^{2}\left\| \Psi (\delta
)(k-k^{0})\right\| ^{2}\leq \left\| \Omega ^{-1/2}\right\| ^{2}\left\| \Psi
(\delta )\right\| ^{2}\left\| k-k^{0}\right\| ^{2}=O(T)\left\|
k-k^{0}\right\| ^{2},
\end{eqnarray*}%
where the last equality follows from Lemma 1 of Perron and Zhu (2005). In
addition, the $O(T)$ term in the above equation is not $o(T)$ and $(XX)>0$.%
\begin{eqnarray*}
\left| (XU)\right|  &=&\left| \theta ^{\prime }(X^{0}-X)^{\prime }\Omega
^{-1/2\prime }M(\lambda )\Omega ^{-1/2}u\right| =\left| (k-k^{0})^{\prime
}\Psi (\delta )^{\prime }\Omega ^{-1/2\prime }M(\lambda )\Omega
^{-1/2}u\right|  \\
&\leq &\left\| k-k^{0}\right\| \left\| \Psi (\delta )^{\prime }\Omega
^{-1/2\prime }M(\lambda )\Omega ^{-1/2}u\right\| =\left\| k-k^{0}\right\|
O_{p}(T^{1/2}).
\end{eqnarray*}%
For $(UU)$, note that%
\begin{eqnarray*}
(UU) &=&u^{\prime }\Omega ^{-1/2\prime }\left[ M(\lambda )-M(\lambda ^{0})%
\right] \Omega ^{-1/2}u \\
&=&u^{\prime }\Omega ^{-1}X^{0}(X^{0\prime }\Omega
^{-1}X^{0})^{-1}X^{0\prime }\Omega ^{-1}u-u^{\prime }\Omega ^{-1}X(X^{\prime
}\Omega ^{-1}X)^{-1}X^{\prime }\Omega ^{-1}u \\
&=&u^{\prime }\Omega ^{-1}(X^{0}-X)(X^{0\prime }\Omega
^{-1}X^{0})^{-1}X^{0\prime }\Omega ^{-1}u \\
&&+u^{\prime }\Omega ^{-1}X(X^{\prime }\Omega ^{-1}X)^{-1}\left( X^{\prime
}\Omega ^{-1}X-X^{0\prime }\Omega ^{-1}X^{0}\right) (X^{0\prime }\Omega
^{-1}X^{0})^{-1}X^{0\prime }\Omega ^{-1}u \\
&&+u^{\prime }\Omega ^{-1}X(X^{\prime }\Omega ^{-1}X)^{-1}(X_{0}-X)^{\prime
}\Omega ^{-1}u \\
&=&\left\| k-k^{0}\right\| O_{p}(T^{-1})\text{.}
\end{eqnarray*}%
These orders of magnitude show that $(XX)$, which is strictly positive, is
the dominant term in (\ref{XX+2XU+UU}). When these terms are evaluated at $k=%
\hat{k}_{r}$, $(XX)$ is still the dominant term and it contradicts to the
fact that $tr[\Sigma ^{-1}(\tilde{U}_{\hat{\lambda}_{r}}^{\prime }\tilde{U}_{%
\hat{\lambda}_{r}}-\tilde{U}_{\lambda ^{0}}^{\prime }\tilde{U}_{\lambda
^{0}})]<0$. The only way to avoid the contradiction is to have $T^{1/2}||%
\hat{k}_{r}-k^{0}||=T^{3/2}(\hat{\lambda}_{r}-\lambda ^{0})=O_{p}(1)$. Since
the rate of convergence is obtained as $T^{3/2}$, the minimization can be
carried out over the set%
\begin{equation*}
\Lambda _{T}=\{\lambda :|\lambda _{ij}-\lambda _{ij}^{0}|\leq MT^{-3/2}\text{
for all }i\text{, }j\text{ and some large constant }M\}\text{.}
\end{equation*}%
On $\Lambda _{T}$, $(UU)$ is asymptotically negligible and can be ignored.
Thus,%
\begin{equation*}
\min_{\lambda \text{ s.t. }R\lambda =r\text{ on }\Lambda _{T}}tr[\Sigma
^{-1}(\tilde{U}_{\lambda }^{\prime }\tilde{U}_{\lambda }-\tilde{U}_{\lambda
^{0}}^{\prime }\tilde{U}_{\lambda ^{0}})]=\min_{\lambda \text{ s.t. }%
R\lambda =r\text{ on }\Lambda _{T}}(XX)+2(XU)+o_{p}(1)\text{.}
\end{equation*}%
Defining $\varphi _{T}=T^{3/2}(\lambda -\lambda ^{0})$ yields that%
\begin{equation*}
(XX)=\varphi _{T}^{\prime }\left[ T^{-1}\Psi (\delta )^{\prime }\Omega
^{-1/2}M(\lambda )\Omega ^{-1/2}\Psi (\delta )\right] \varphi _{T},
\end{equation*}%
and%
\begin{equation*}
(XU)=\varphi _{T}^{\prime }\left[ T^{-1/2}\Psi (\delta )^{\prime }\Omega
^{-1/2\prime }M(\lambda )\Omega ^{-1/2}u\right] \text{.}
\end{equation*}%
Define a $Tn\times m$ matrix $Z=T^{-1/2}M(\lambda ^{0})\Omega ^{-1/2}\Psi
(\delta )$, a $Tn\times 1$ vector $z=M(\lambda ^{0})\Omega ^{-1/2}u$ and $%
Q=Z(Z^{\prime }Z)^{-1}R_{\perp }^{\prime }$ where the $(m-q)\times m$ matrix 
$R_{\perp }$ is such that $R(Z^{\prime }Z)^{-1}R_{\perp }^{\prime }=0$ for a 
$q\times m$ matrix of restrictions $R$. Then,%
\begin{eqnarray*}
\min_{\lambda \text{ s.t. }R\lambda =r\text{ on }\Lambda _{T}}(XX)+2(XU)
&=&\min_{\varphi _{T}\text{ s.t. }R\varphi _{T}=0\text{ on }\Lambda _{T}}%
\left[ Z\varphi _{T}+z\right] ^{\prime }\left[ Z\varphi _{T}+z\right]
-z^{\prime }z+o_{p}(1) \\
&=&z\left[ I-Q(Q^{\prime }Q)^{-1}Q^{\prime }\right] z-z^{\prime }z+o_{p}(1)
\\
&=&-z^{\prime }Q(Q^{\prime }Q)^{-1}Q^{\prime }z+o_{p}(1).
\end{eqnarray*}%
The second term in (\ref{LR decomposition}) is basically the same as the
first except that there is no restriction on the break dates, so that $Q=Z$.
It follows that%
\begin{equation*}
T~tr\left[ \hat{\Sigma}^{-1}(\hat{\lambda})\left( \hat{\Sigma}(\hat{\lambda}%
)-\hat{\Sigma}(\lambda ^{0})\right) \right] =-z^{\prime }Z(Z^{\prime
}Z)^{-1}Z^{\prime }z+o_{p}(1)\text{,}
\end{equation*}%
and%
\begin{eqnarray*}
LR &=&-z^{\prime }Q(Q^{\prime }Q)^{-1}Q^{\prime }z+z^{\prime }Z(Z^{\prime
}Z)^{-1}Z^{\prime }z+o_{p}(1) \\
&=&z^{\prime }Z(Z^{\prime }Z)^{-1}R\left( R^{\prime }(Z^{\prime
}Z)^{-1}R\right) ^{-1}R^{\prime }(Z^{\prime }Z)^{-1}Z^{\prime }z+o_{p}(1)%
\text{.}
\end{eqnarray*}%
Note that%
\begin{eqnarray*}
Z^{\prime }Z &=&\frac{1}{T}\Psi (\delta )^{\prime }\Omega ^{-1/2\prime
}M(\lambda ^{0})\Omega ^{-1/2}\Psi (\delta ) \\
&=&\frac{1}{T}\Psi (\delta )^{\prime }\Omega ^{-1}\Psi (\delta )-\frac{1}{T}%
\Psi (\delta )^{\prime }\Omega ^{-1}X^{0}S_{T}\left( \frac{1}{T}%
S_{T}X^{0\prime }\Omega ^{-1}X^{0}S_{T}\right) ^{-1}\frac{1}{T}%
S_{T}X^{0\prime }\Omega ^{-1}\Psi (\delta ) \\
&\rightarrow &D_{\delta }\left[ Q_{GG}-Q_{GF}Q_{FF}^{-1}Q_{FG}\right]
D_{\delta }=\Xi _{1}\text{,}
\end{eqnarray*}%
where%
\begin{equation*}
S_{T}=diag\left\{ 
\begin{bmatrix}
1 &  &  \\ 
& T^{-1} &  \\ 
&  & T^{-1}I_{m_{1}}%
\end{bmatrix}%
,\dots ,%
\begin{bmatrix}
1 &  &  \\ 
& T^{-1} &  \\ 
&  & T^{-1}I_{m_{n}}%
\end{bmatrix}%
\right\} \text{.}
\end{equation*}%
For $Z^{\prime }z$, let $\epsilon =\Omega ^{-1}u$, which means that $%
\epsilon _{t}=(\epsilon _{1t},...,\epsilon _{nt})^{\prime }=\Sigma
^{-1}(u_{1t},...,u_{nt})^{\prime }$ and the partial sums of $\epsilon _{t}$
obeys a functional central limit theorem from Assumption \ref{A3}, i.e., $%
T^{-1/2}\sum_{t=1}^{[Tr]}\epsilon _{t}\Rightarrow \bar{B}(r)=\Sigma
^{-1}\Psi ^{1/2}W(r)$. Denote the $i^{th}$ element of $\bar{B}(r)$ by $\bar{B%
}_{i}(r)$. We can then express the limit of $Z^{\prime }z$ as the follows:%
\begin{eqnarray*}
Z^{\prime }z &=&T^{-1/2}\Psi (\delta )^{\prime }\Omega ^{-1/2\prime
}M(\lambda ^{0})\Omega ^{-1/2}u \\
&=&T^{-1/2}\Psi (\delta )^{\prime }\Omega ^{-1}u-T^{-1}\Psi (\delta
)^{\prime }\Omega ^{-1}X^{0}S_{T}\left( T^{-1}S_{T}X^{0\prime }\Omega
^{-1}X^{0}S_{T}\right) ^{-1}T^{-1/2}S_{T}X^{0\prime }\Omega ^{-1}u \\
&=&%
\begin{bmatrix}
-T^{-1}\Psi (\delta )^{\prime }\Omega ^{-1}X^{0}S_{T}\left( \frac{1}{T}%
S_{T}X^{0\prime }\Omega ^{-1}X^{0}S_{T}\right) ^{-1} & I%
\end{bmatrix}%
\begin{bmatrix}
T^{-1/2}S_{T}X^{0\prime }\epsilon  \\ 
T^{-1/2}\Psi (\delta )^{\prime }\epsilon 
\end{bmatrix}
\\
&&\overset{d}{\rightarrow }%
\begin{bmatrix}
-D_{\delta }Q_{GF}Q_{FF}^{-1} & I_{m}%
\end{bmatrix}%
\times  \\
&&\left( 
\begin{array}{cccccc}
\int f_{1}(r)^{\prime }d\bar{B}_{1}(r) & \cdots  & \int f_{n}(r)^{\prime }d%
\bar{B}_{n}(r) & \int g_{1}(r)^{\prime }d\bar{B}_{1}(r) & \cdots  & \int
g_{n}(r)^{\prime }d\bar{B}_{n}(r)%
\end{array}%
\right) ^{\prime },
\end{eqnarray*}%
where the covariance of the vector of stochastic integrals in the above
expression is given by%
\begin{equation*}
\left( 
\begin{array}{cc}
\Gamma _{FF} & \Gamma _{FG}D_{\delta } \\ 
D_{\delta }\Gamma _{GF} & D_{\delta }\Gamma _{GG}D_{\delta }%
\end{array}%
\right) \text{.}
\end{equation*}%
Thus, $Z^{\prime }z\overset{d}{\rightarrow }N\left( 0,\Xi _{0}\right) $.
Theorem \ref{Theorem 1} assumes that $\Sigma =\Psi $, hence $\Xi _{0}=\Xi
_{1}$. Then, it follows that $LR\overset{d}{\rightarrow }\chi _{q}^{2}$. For
Theorem \ref{Theorem 2}, $T^{3/2}(\hat{\lambda}-\lambda ^{0})=\hat{\varphi}%
_{T}+o_{p}(1)=-(Z^{\prime }Z)^{-1}Z^{\prime }z\overset{d}{\rightarrow }%
N(0,\Xi _{1}^{-1}\Xi _{0}\Xi _{1}^{-1\prime })$. \vspace{0.1in}

\noindent \textbf{Proof of Corollary \ref{Corollary 1}: }Similarly to (\ref%
{XX+2XU+UU}), we have the following decomposition%
\begin{eqnarray*}
SSR_{i}(\lambda _{i})-SSR_{i}(\lambda _{i}^{0}) &=&\theta _{i}^{\prime
}(X(k_{i}^{0})-X(k_{i}))^{\prime }M_{\lambda
_{i}}(X(k_{i}^{0})-X(k_{i}))\theta _{i} \\
&&+2\theta _{i}^{\prime }(X(k_{i}^{0})-X(k_{i}))^{\prime }M_{\lambda
_{i}}u_{i} \\
&&+u_{i}^{\prime }(M_{\lambda _{i}}-M_{\lambda _{i}^{0}})u_{i} \\
&\equiv &(XX)_{i}+2(XU)_{i}+(UU)_{i}.
\end{eqnarray*}%
The results in Perron and Zhu (2005) directly applies and $\tilde{\lambda}%
_{i}$ is consistent at rate $T^{3/2}$. Thus, we can again focus on $T^{-3/2}$
neighborhood of $\lambda _{i}^{0}$, say $\Lambda _{iT}$, and%
\begin{equation*}
\tilde{\lambda}_{i}=\arg \min_{\lambda _{i}\in \Lambda _{iT}}SSR_{i}(\lambda
_{i})-SSR_{i}(\lambda _{i}^{0})=\arg \min_{\lambda _{i}\in \Lambda
_{iT}}(XX)_{i}+2(XU)_{i}+o_{p}(1)\text{.}
\end{equation*}%
Define $\varphi _{i}=T^{3/2}(\lambda _{i}-\lambda _{i}^{0})$ and $\Psi
_{i}(\delta _{i})=\tilde{\iota}(k_{i})D_{\delta _{i}}$. It follows that%
\begin{eqnarray*}
T^{3/2}(\tilde{\lambda}_{i}-\lambda _{i}^{0}) &=&\arg \min_{\varphi _{i}\in
\Lambda _{iT}}\varphi _{i}^{\prime }\left[ T^{-1}\Psi _{i}(\delta
_{i})^{\prime }M_{\lambda _{i}}\Psi _{i}(\delta _{i})\right] \varphi
_{i}+2\varphi _{i}^{\prime }\left[ T^{-1/2}\Psi _{i}(\delta _{i})^{\prime
}M_{\lambda _{i}}u_{i}\right] +o_{p}(1) \\
&=&-\left[ T^{-1}\Psi _{i}(\delta _{i})^{\prime }M_{\lambda _{i}^{0}}\Psi
_{i}(\delta _{i})\right] ^{-1}\left[ T^{-1/2}\Psi _{i}(\delta _{i})^{\prime
}M_{\lambda _{i}^{0}}u_{i}\right] +o_{p}(1)\text{.}
\end{eqnarray*}%
Note that 
\begin{equation*}
T^{-1}\Psi _{i}(\delta _{i})^{\prime }M_{\lambda _{i}^{0}}\Psi _{i}(\delta
_{i})\rightarrow D_{\delta _{i}}\left[ \int p_{i}(r)p_{i}^{\prime }(r)dr%
\right] D_{\delta _{i}}
\end{equation*}%
and%
\begin{equation*}
T^{-1/2}\Psi _{i}(\delta _{i})^{\prime }M_{\lambda _{i}^{0}}u_{i}\Rightarrow
D_{\delta _{i}}\left[ \int p_{i}(r)dB_{i}(r)\right] ,
\end{equation*}%
where $B_{i}(r)$ is the $i^{th}$ element of $B(r)=\Psi ^{1/2}W(r)$ in
Assumption \ref{A3}. Furthermore, the weak convergence holds jointly in $i$
under Assumption \ref{A3}. Therefore,%
\begin{equation*}
T^{3/2}%
\begin{pmatrix}
\tilde{\lambda}_{1}-\lambda _{1}^{0} \\ 
\vdots \\ 
\tilde{\lambda}_{n}-\lambda _{n}^{0}%
\end{pmatrix}%
\overset{d}{\rightarrow }N\left( 0,%
\begin{pmatrix}
\psi _{11}D_{\delta _{1}}^{-1}P_{11}D_{\delta _{1}}^{-1} & \dots & \psi
_{1n}D_{\delta _{1}}^{-1}P_{1n}D_{\delta _{n}}^{-1} \\ 
\vdots & \ddots & \vdots \\ 
\psi _{n1}D_{\delta _{n}}^{-1}P_{n1}D_{\delta _{1}}^{-1} & \dots & \psi
_{nn}D_{\delta _{n}}^{-1}P_{nn}D_{\delta _{n}}^{-1}%
\end{pmatrix}%
\right) ,
\end{equation*}%
where $P_{ij}=(\int p_{i}(r)p_{i}^{\prime }(r)dr)^{-1}\int
p_{i}(r)p_{j}^{\prime }(r)dr(\int p_{j}(r)p_{j}^{\prime }(r)dr)^{-1}$. 
\vspace{0.1in}

\noindent \textbf{Proof of Theorem \ref{Theorem 3}:} We use $\hat{\Sigma}%
_{(i)}(k,h)$ and $\hat{\Sigma}_{(i)}(\lambda ,\nu )$ interchangeably. As in (%
\ref{LR decomposition}), 
\begin{equation*}
LR=T~tr\left[ \hat{\Sigma}_{(i)}^{-1}(\hat{\lambda},\hat{\nu})\left( \hat{%
\Sigma}(\hat{\lambda})-\hat{\Sigma}_{(i)}(\hat{\lambda},\hat{\nu})\right) %
\right] +o_{p}(1)\text{,}
\end{equation*}%
where $\hat{\nu}=\arg \min_{\nu \in C_{T}^{(i)}}\log \left| \hat{\Sigma}%
_{(i)}(\hat{\lambda},\nu )\right| $. Define the projection matrix $%
P_{(i)}(\lambda ,\nu )$ as%
\begin{equation*}
P_{(i)}(\lambda ,\nu )=M(\lambda )\Omega ^{-1/2}a_{i}\left( a_{i}^{\prime
}\Omega ^{-1/2\prime }M(\lambda )\Omega ^{-1/2}a_{i}\right)
^{-1}a_{i}^{\prime }\Omega ^{-1/2\prime }M(\lambda )\text{,}
\end{equation*}%
where $a_{i}=a_{i}([T\nu ])$ for simplicity and $M(\lambda )=I-\Omega
^{-1/2}X(X^{\prime }\Omega ^{-1}X)^{-1}X^{\prime }\Omega ^{-1/2}$. Then, it
follows that%
\begin{equation*}
LR=\sup\nolimits_{\nu \in C_{T}^{(i)}}y^{\prime }\Omega ^{-1/2\prime }M(\hat{%
\lambda})P_{(i)}(\hat{\lambda},\nu )M(\hat{\lambda})\Omega ^{-1/2}y+o_{p}(1)%
\text{.}
\end{equation*}%
Since%
\begin{eqnarray*}
M(\lambda )\Omega ^{-1/2}y &=&M(\lambda )\Omega ^{-1/2}(X^{0}-X)\theta
+M(\lambda )\Omega ^{-1/2}u \\
&=&M(\lambda )\Omega ^{-1/2}\Psi (\delta )(k-k^{0})+M(\lambda )\Omega
^{-1/2}u,
\end{eqnarray*}%
we can write%
\begin{equation*}
LR=\sup_{\nu \in C_{T}^{(i)}}\left[ 
\begin{array}{c}
(\hat{k}-k^{0})^{\prime }\Psi (\delta )^{\prime }\Omega ^{-1/2\prime }M(\hat{%
\lambda})P_{(i)}(\hat{\lambda},\nu )M(\hat{\lambda})\Omega ^{-1/2}\Psi
(\delta )(\hat{k}-k^{0}) \\ 
+2(\hat{k}-k^{0})^{\prime }\Psi (\delta )^{\prime }\Omega ^{-1/2\prime }M(%
\hat{\lambda})P_{(i)}(\hat{\lambda},\nu )M(\hat{\lambda})\Omega ^{-1/2}u \\ 
+u^{\prime }\Omega ^{-1/2\prime }M(\hat{\lambda})P_{(i)}(\hat{\lambda},\nu
)M(\hat{\lambda})\Omega ^{-1/2}u%
\end{array}%
\right] +o_{p}(1)\text{.}
\end{equation*}%
From the proof of Theorem \ref{Theorem 1}, recall that $\sqrt{T}(\hat{k}%
-k^{0})=-(Z^{\prime }Z)^{-1}Z^{\prime }z+o_{p}(1)$ where $%
Z=T^{-1/2}M(\lambda ^{0})\Omega ^{-1/2}\Psi (\delta )$ and $z=M(\lambda
^{0})\Omega ^{-1/2}u$. Then,%
\begin{equation*}
LR=\sup\nolimits_{\nu \in C^{(i)}}\left[ 
\begin{array}{c}
z^{\prime }\left( I-Z(Z^{\prime }Z)^{-1}Z^{\prime }\right) P_{(i)}(\lambda
^{0},\nu )\left( I-Z(Z^{\prime }Z)^{-1}Z^{\prime }\right) z%
\end{array}%
\right] +o_{p}(1)\text{.}
\end{equation*}%
To further analyze this expression, we write%
\begin{eqnarray}
&&z^{\prime }\left( I-Z(Z^{\prime }Z)^{-1}Z^{\prime }\right) P_{(i)}(\lambda
^{0},\nu )\left( I-Z(Z^{\prime }Z)^{-1}Z^{\prime }\right) z  \label{zPz} \\
&=&\frac{(T^{-3/2}a_{i}^{\prime }\Omega ^{-1/2\prime }M(\lambda ^{0})\left(
I-Z(Z^{\prime }Z)^{-1}Z^{\prime }\right) z)^{2}}{T^{-3}a_{i}^{\prime }\Omega
^{-1/2\prime }M(\lambda ^{0})\Omega ^{-1/2}a_{i}}.  \notag
\end{eqnarray}%
The weak convergence result for the numerator of (\ref{zPz}) follows from
Assumption \ref{A3} and the continuous mapping theorem. To see this,
consider the following decomposition.%
\begin{equation*}
T^{-3/2}a_{i}^{\prime }\Omega ^{-1/2\prime }M(\lambda ^{0})\left(
I-Z(Z^{\prime }Z)^{-1}Z^{\prime }\right) z=A_{1}(\nu )A_{2}(\nu )A_{3}(\nu )%
\text{, say,}
\end{equation*}%
with%
\begin{eqnarray*}
A_{1}(\nu ) &=&\left( 1,-\left( T^{-3/2}a_{i}^{\prime }\Omega ^{-1/2\prime
}M(\lambda ^{0})Z\right) (Z^{\prime }Z)^{-1}\right) \\
A_{2}(\nu ) &=&\left( 
\begin{array}{ccc}
1 & -T^{-2}a_{i}^{\prime }\Omega ^{-1}X^{0}S_{T}\left( \frac{1}{T}%
S_{T}X^{0\prime }\Omega ^{-1}X^{0}S_{T}\right) ^{-1} & 0 \\ 
0 & -T^{-1}\Psi (\delta )^{\prime }\Omega ^{-1}X^{0}S_{T}\left( \frac{1}{T}%
S_{T}X^{0\prime }\Omega ^{-1}X^{0}S_{T}\right) ^{-1} & I_{m}%
\end{array}%
\right) \\
A_{3}(\nu ) &=&\left( 
\begin{array}{ccc}
T^{-3/2}a_{i} & T^{-1/2}(S_{T}X^{0\prime }) & T^{-1/2}\Psi (\delta )%
\end{array}%
\right) ^{\prime }\Omega ^{-1}u\text{.}
\end{eqnarray*}%
For $A_{1}$, note that $Z^{\prime }Z\overset{p}{\rightarrow }\Xi _{1}$ from
the proof of Theorem \ref{Theorem 1}\ and%
\begin{eqnarray*}
T^{-3/2}a_{i}^{\prime }\Omega ^{-1/2\prime }M(\lambda ^{0})Z
&=&T^{-2}a_{i}^{\prime }\Omega ^{-1/2\prime }M(\lambda ^{0})\Omega
^{-1/2}\Psi (\delta ) \\
&=&T^{-2}a_{i}^{\prime }\Omega ^{-1}\Psi (\delta ) \\
&&-T^{-2}a_{i}^{\prime }\Omega ^{-1}X^{0}S_{T}\left( \frac{1}{T}%
S_{T}X^{0\prime }\Omega ^{-1}X^{0}S_{T}\right) ^{-1}T^{-1}S_{T}X^{0\prime
}\Omega ^{-1}\Psi (\delta ) \\
&\rightarrow &[Q_{BG}^{(i)}(\nu )-Q_{BF}^{(i)}(\nu
)Q_{FF}^{-1}Q_{FG}]D_{\delta }\text{,}
\end{eqnarray*}%
uniformly in $\nu $. Thus,%
\begin{equation}
A_{1}(\nu )\rightarrow \left( 1,-[Q_{BG}^{(i)}(\nu )-Q_{BF}^{(i)}(\nu
)Q_{FF}^{-1}Q_{FG}]D_{\delta }\Xi _{1}^{-1}\right) =\left( 1,-\varsigma
_{1}^{\prime }(\nu )\Xi _{1}^{-1}\right) \text{,}  \label{A1v}
\end{equation}%
uniformly in $\nu $. Similarly, $A_{2}$ is such that%
\begin{equation}
A_{2}(\nu )\rightarrow \left( 
\begin{array}{ccc}
1 & -Q_{BF}^{(i)}(\nu )Q_{FF}^{-1} & 0 \\ 
0 & -D_{\delta }Q_{GF}Q_{FF}^{-1} & I_{m}%
\end{array}%
\right) \text{,}  \label{A2v}
\end{equation}%
uniformly in $\nu $. For $A_{3}$, let $\bar{B}_{i}(r)$ be as defined in the
proof of Theorem \ref{Theorem 1}. Then, 
\begin{equation}
A_{3}(\nu )=\left( 
\begin{array}{c}
T^{-3/2}a_{i}^{\prime }\epsilon \\ 
T^{-1/2}S_{T}X^{0\prime }\epsilon \\ 
T^{-1/2}\Psi (\delta )^{\prime }\epsilon%
\end{array}%
\right) \Rightarrow \left( 
\begin{array}{ccc}
1 &  & 0 \\ 
& I_{m+2n} &  \\ 
0 &  & D_{\delta }%
\end{array}%
\right) \left( 
\begin{array}{c}
\int b(r,\nu )d\bar{B}_{i}(r) \\ 
\int f_{1}(r)d\bar{B}_{1}(r) \\ 
\vdots \\ 
\int f_{n}(r)d\bar{B}_{n}(r) \\ 
\int g_{1}(r)d\bar{B}_{1}(r) \\ 
\vdots \\ 
\int g_{n}(r)d\bar{B}_{n}(r)%
\end{array}%
\right) ,  \label{A3v}
\end{equation}%
over $\nu \in (0,1)$. Thus, $A_{3}$ is a Gaussian process with covariance
matrix given by%
\begin{equation*}
\left( 
\begin{array}{ccc}
\Gamma _{BB}^{(i)}(\nu ) & \Gamma _{BF}^{(i)}(\nu ) & \Gamma _{BG}^{(i)}(\nu
)D_{\delta } \\ 
\Gamma _{FB}^{(i)}(\nu ) & \Gamma _{FF} & \Gamma _{FG}D_{\delta } \\ 
D_{\delta }\Gamma _{GB}^{(i)}(\nu ) & D_{\delta }\Gamma _{GF} & D_{\delta
}\Gamma _{GG}D_{\delta }%
\end{array}%
\right) \text{.}
\end{equation*}%
Therefore, combining (\ref{A1v}), (\ref{A2v}) and (\ref{A3v}) yields the
weak convergence result%
\begin{equation}
T^{-3/2}a_{i}^{\prime }\Omega ^{-1/2\prime }M(\lambda ^{0})[I-Z(Z^{\prime
}Z)^{-1}Z^{\prime }]z\Rightarrow \eta _{(i)}(\nu ),  \label{numer}
\end{equation}%
over $\nu \in (0,1)$. Lastly, the denominator in (\ref{zPz}) is such that%
\begin{eqnarray}
&&T^{-3}a_{i}^{\prime }\Omega ^{-1/2\prime }M(\lambda ^{0})\Omega
^{-1/2}a_{i}  \label{deno} \\
&=&T^{-3}a_{i}^{\prime }\Omega ^{-1}a_{i}-T^{-2}a_{i}^{\prime }\Omega
^{-1}X^{0}S_{T}\left( T^{-1}S_{T}X^{0\prime }\Omega ^{-1}X^{0}S_{T}\right)
^{-1}T^{-2}S_{T}X^{0\prime }\Omega ^{-1}a_{i}  \notag \\
&\rightarrow &Q_{BB}^{(i)}(\nu )-Q_{BF}^{(i)}(\nu
)Q_{FF}^{-1}Q_{FB}^{(i)}(\nu )=\xi _{1}^{(i)}(\nu )\text{,}  \notag
\end{eqnarray}%
uniformly in $\nu $. The result in part (i) of the theorem follows from (\ref%
{numer}) and (\ref{deno}). The result in part (ii) also follows since the
above result holds jointly in $i$. \pagebreak

\newpage
\setcounter{page}{1}\renewcommand{\thepage}{T-\arabic{page}}

\begin{center}
Table 1. Probabilities of rejecting the null hypothesis, LR test

\bigskip

(a) $\delta _{i}=0.5$

\begingroup \scalefont{0.8}\renewcommand{\arraystretch}{1.0}

\begin{tabular}{cc|cccc|cccc}
\hline
&  & \multicolumn{4}{|c}{Asymptotic} & \multicolumn{4}{|c}{Bootstrap} \\ 
\hline
$\alpha $ & $\rho $ & (1) & (2) & (3) & (4) & (1) & (2) & (3) & (4) \\ \hline
0.0 & -0.5 & 0.05 & 0.98 & 0.06 & 0.95 & 0.04 & 0.97 & 0.05 & 0.94 \\ 
0.0 & 0.0 & 0.07 & 1.00 & 0.07 & 0.98 & 0.06 & 0.98 & 0.05 & 0.97 \\ 
0.0 & 0.5 & 0.05 & 1.00 & 0.05 & 1.00 & 0.05 & 1.00 & 0.04 & 1.00 \\ 
0.3 & -0.5 & 0.17 & 0.99 & 0.21 & 0.99 & 0.05 & 0.98 & 0.06 & 0.95 \\ 
0.3 & 0.0 & 0.17 & 1.00 & 0.18 & 1.00 & 0.05 & 0.99 & 0.06 & 0.97 \\ 
0.3 & 0.5 & 0.18 & 1.00 & 0.15 & 1.00 & 0.07 & 1.00 & 0.05 & 1.00 \\ 
0.7 & -0.5 & 0.49 & 1.00 & 0.52 & 1.00 & 0.08 & 0.97 & 0.08 & 0.96 \\ 
0.7 & 0.0 & 0.50 & 1.00 & 0.40 & 1.00 & 0.04 & 0.99 & 0.07 & 0.98 \\ 
0.7 & 0.5 & 0.47 & 1.00 & 0.33 & 1.00 & 0.08 & 1.00 & 0.05 & 1.00 \\ \hline
\end{tabular}

\endgroup

\bigskip

(b) $\delta _{i}=1.0$

\begingroup \scalefont{0.8}\renewcommand{\arraystretch}{1.0}

\begin{tabular}{cc|cccc|cccc}
\hline
&  & \multicolumn{4}{|c}{Asymptotic} & \multicolumn{4}{|c}{Bootstrap} \\ 
\hline
$\alpha $ & $\rho $ & (1) & (2) & (3) & (4) & (1) & (2) & (3) & (4) \\ \hline
0.0 & -0.5 & 0.04 & 1.00 & 0.09 & 1.00 & 0.06 & 1.00 & 0.04 & 1.00 \\ 
0.0 & 0.0 & 0.04 & 1.00 & 0.07 & 1.00 & 0.05 & 1.00 & 0.06 & 1.00 \\ 
0.0 & 0.5 & 0.05 & 1.00 & 0.05 & 1.00 & 0.07 & 1.00 & 0.06 & 1.00 \\ 
0.3 & -0.5 & 0.10 & 1.00 & 0.14 & 1.00 & 0.06 & 1.00 & 0.06 & 1.00 \\ 
0.3 & 0.0 & 0.11 & 1.00 & 0.14 & 1.00 & 0.04 & 1.00 & 0.05 & 1.00 \\ 
0.3 & 0.5 & 0.12 & 1.00 & 0.10 & 1.00 & 0.04 & 1.00 & 0.04 & 1.00 \\ 
0.7 & -0.5 & 0.14 & 1.00 & 0.17 & 1.00 & 0.07 & 1.00 & 0.05 & 1.00 \\ 
0.7 & 0.0 & 0.20 & 1.00 & 0.22 & 1.00 & 0.06 & 1.00 & 0.08 & 1.00 \\ 
0.7 & 0.5 & 0.14 & 1.00 & 0.11 & 1.00 & 0.05 & 1.00 & 0.05 & 1.00 \\ \hline
\end{tabular}

\endgroup

\bigskip

(c) $\delta _{i}=1.5$

\begingroup \scalefont{0.8}\renewcommand{\arraystretch}{1.0}

\begin{tabular}{cc|cccc|cccc}
\hline
&  & \multicolumn{4}{|c}{Asymptotic} & \multicolumn{4}{|c}{Bootstrap} \\ 
\hline
$\alpha $ & $\rho $ & (1) & (2) & (3) & (4) & (1) & (2) & (3) & (4) \\ \hline
0.0 & -0.5 & 0.01 & 1.00 & 0.02 & 1.00 & 0.06 & 1.00 & 0.06 & 1.00 \\ 
0.0 & 0.0 & 0.02 & 1.00 & 0.03 & 1.00 & 0.05 & 1.00 & 0.05 & 1.00 \\ 
0.0 & 0.5 & 0.01 & 1.00 & 0.02 & 1.00 & 0.07 & 1.00 & 0.05 & 1.00 \\ 
0.3 & -0.5 & 0.03 & 1.00 & 0.04 & 1.00 & 0.07 & 1.00 & 0.07 & 1.00 \\ 
0.3 & 0.0 & 0.03 & 1.00 & 0.04 & 1.00 & 0.03 & 1.00 & 0.04 & 1.00 \\ 
0.3 & 0.5 & 0.02 & 1.00 & 0.02 & 1.00 & 0.04 & 1.00 & 0.03 & 1.00 \\ 
0.7 & -0.5 & 0.02 & 1.00 & 0.02 & 1.00 & 0.03 & 1.00 & 0.03 & 1.00 \\ 
0.7 & 0.0 & 0.04 & 1.00 & 0.05 & 1.00 & 0.05 & 1.00 & 0.07 & 1.00 \\ 
0.7 & 0.5 & 0.03 & 1.00 & 0.02 & 1.00 & 0.04 & 1.00 & 0.02 & 1.00 \\ \hline
\end{tabular}

\endgroup

\bigskip

\parbox{16cm}
{\begin{small} Note: $T=100$. The true break dates are $T/2$. The null hypotheses are as follows: (1) $R=I$ and $r=(0.5,0.5)'$, (2) $R=I$ and $r=(0.525,0.475)'$, (3) $R=[1,-1]$ and $r=0$ and (4) $R=[1,-1]$ and $r=0.05$. Hence, the null rejection probabilities for (1) and (3) stand for the finite sample sizes while those for (2) and (4) are powers.
\end{small}}
\end{center}

\newpage

\begin{center}
Table 2. Probabilities of rejecting the null hypothesis, GLS-Wald test

\bigskip

(a) $\delta _{i}=0.5$

\begingroup \scalefont{0.8}\renewcommand{\arraystretch}{1.0}

\begin{tabular}{cc|cccc|cccc}
\hline
&  & \multicolumn{4}{|c}{Asymptotic} & \multicolumn{4}{|c}{Bootstrap} \\ 
\hline
$\alpha $ & $\rho $ & (1) & (2) & (3) & (4) & (1) & (2) & (3) & (4) \\ \hline
0.0 & -0.5 & 0.13 & 0.99 & 0.08 & 0.96 & 0.05 & 0.94 & 0.04 & 0.90 \\ 
0.0 & 0.0 & 0.13 & 1.00 & 0.08 & 0.98 & 0.07 & 0.98 & 0.04 & 0.97 \\ 
0.0 & 0.5 & 0.15 & 1.00 & 0.07 & 1.00 & 0.05 & 1.00 & 0.04 & 1.00 \\ 
0.3 & -0.5 & 0.14 & 0.99 & 0.14 & 0.97 & 0.06 & 0.93 & 0.05 & 0.88 \\ 
0.3 & 0.0 & 0.15 & 1.00 & 0.15 & 0.99 & 0.05 & 0.98 & 0.04 & 0.97 \\ 
0.3 & 0.5 & 0.17 & 1.00 & 0.09 & 1.00 & 0.06 & 1.00 & 0.05 & 1.00 \\ 
0.7 & -0.5 & 0.43 & 1.00 & 0.22 & 0.99 & 0.06 & 0.67 & 0.05 & 0.82 \\ 
0.7 & 0.0 & 0.37 & 1.00 & 0.21 & 1.00 & 0.03 & 0.87 & 0.05 & 0.92 \\ 
0.7 & 0.5 & 0.40 & 1.00 & 0.27 & 1.00 & 0.06 & 0.99 & 0.05 & 1.00 \\ \hline
\end{tabular}

\endgroup

\bigskip

(b) $\delta _{i}=1.0$

\begingroup \scalefont{0.8}\renewcommand{\arraystretch}{1.0}

\begin{tabular}{cc|cccc|cccc}
\hline
&  & \multicolumn{4}{|c}{Asymptotic} & \multicolumn{4}{|c}{Bootstrap} \\ 
\hline
$\alpha $ & $\rho $ & (1) & (2) & (3) & (4) & (1) & (2) & (3) & (4) \\ \hline
0.0 & -0.5 & 0.32 & 1.00 & 0.09 & 1.00 & 0.03 & 1.00 & 0.06 & 1.00 \\ 
0.0 & 0.0 & 0.29 & 1.00 & 0.09 & 1.00 & 0.06 & 1.00 & 0.03 & 1.00 \\ 
0.0 & 0.5 & 0.33 & 1.00 & 0.23 & 1.00 & 0.03 & 1.00 & 0.05 & 1.00 \\ 
0.3 & -0.5 & 0.28 & 1.00 & 0.13 & 1.00 & 0.03 & 1.00 & 0.06 & 1.00 \\ 
0.3 & 0.0 & 0.35 & 1.00 & 0.23 & 1.00 & 0.04 & 1.00 & 0.03 & 1.00 \\ 
0.3 & 0.5 & 0.34 & 1.00 & 0.23 & 1.00 & 0.04 & 1.00 & 0.04 & 1.00 \\ 
0.7 & -0.5 & 0.21 & 1.00 & 0.18 & 1.00 & 0.06 & 0.90 & 0.07 & 1.00 \\ 
0.7 & 0.0 & 0.30 & 1.00 & 0.26 & 1.00 & 0.05 & 0.99 & 0.06 & 1.00 \\ 
0.7 & 0.5 & 0.22 & 1.00 & 0.15 & 1.00 & 0.03 & 1.00 & 0.07 & 1.00 \\ \hline
\end{tabular}

\endgroup

\bigskip

(c) $\delta _{i}=1.5$

\begingroup \scalefont{0.8}\renewcommand{\arraystretch}{1.0}

\begin{tabular}{cc|cccc|cccc}
\hline
&  & \multicolumn{4}{|c}{Asymptotic} & \multicolumn{4}{|c}{Bootstrap} \\ 
\hline
$\alpha $ & $\rho $ & (1) & (2) & (3) & (4) & (1) & (2) & (3) & (4) \\ \hline
0.0 & -0.5 & 0.08 & 1.00 & 0.08 & 1.00 & 0.08 & 1.00 & 0.05 & 1.00 \\ 
0.0 & 0.0 & 0.13 & 1.00 & 0.13 & 1.00 & 0.04 & 1.00 & 0.06 & 1.00 \\ 
0.0 & 0.5 & 0.09 & 1.00 & 0.06 & 1.00 & 0.08 & 1.00 & 0.06 & 1.00 \\ 
0.3 & -0.5 & 0.07 & 1.00 & 0.07 & 1.00 & 0.07 & 1.00 & 0.07 & 1.00 \\ 
0.3 & 0.0 & 0.10 & 1.00 & 0.09 & 1.00 & 0.06 & 1.00 & 0.05 & 1.00 \\ 
0.3 & 0.5 & 0.07 & 1.00 & 0.05 & 1.00 & 0.06 & 1.00 & 0.05 & 1.00 \\ 
0.7 & -0.5 & 0.03 & 1.00 & 0.03 & 1.00 & 0.03 & 0.78 & 0.03 & 1.00 \\ 
0.7 & 0.0 & 0.07 & 1.00 & 0.07 & 1.00 & 0.07 & 0.99 & 0.07 & 1.00 \\ 
0.7 & 0.5 & 0.04 & 1.00 & 0.02 & 1.00 & 0.04 & 1.00 & 0.02 & 1.00 \\ \hline
\end{tabular}

\endgroup

\bigskip

\parbox{16cm}
{\begin{small} Note: $T=100$. The true break dates are $T/2$. The null hypotheses are as follows: (1) $R=I$ and $r=(0.5,0.5)'$, (2) $R=I$ and $r=(0.525,0.475)'$, (3) $R=[1,-1]$ and $r=0$ and (4) $R=[1,-1]$ and $r=0.05$. Hence, the null rejection probabilities for (1) and (3) stand for the finite sample sizes while those for (2) and (4) are powers.
\end{small}}
\end{center}

\newpage

\begin{center}
Table 3. Probabilities of rejecting the null hypothesis, OLS-Wald test

\bigskip

(a) $\delta _{i}=0.5$

\begingroup \scalefont{0.8}\renewcommand{\arraystretch}{1.0}

\begin{tabular}{cc|cccc|cccc}
\hline
&  & \multicolumn{4}{|c}{Asymptotic} & \multicolumn{4}{|c}{Bootstrap} \\ 
\hline
$\alpha $ & $\rho $ & (1) & (2) & (3) & (4) & (1) & (2) & (3) & (4) \\ \hline
0.0 & -0.5 & 0.16 & 0.99 & 0.09 & 0.96 & 0.04 & 0.94 & 0.05 & 0.90 \\ 
0.0 & 0.0 & 0.11 & 1.00 & 0.08 & 0.98 & 0.05 & 0.99 & 0.04 & 0.96 \\ 
0.0 & 0.5 & 0.15 & 1.00 & 0.11 & 1.00 & 0.05 & 1.00 & 0.05 & 1.00 \\ 
0.3 & -0.5 & 0.15 & 0.99 & 0.13 & 0.97 & 0.06 & 0.94 & 0.05 & 0.91 \\ 
0.3 & 0.0 & 0.15 & 1.00 & 0.15 & 0.99 & 0.05 & 0.98 & 0.06 & 0.97 \\ 
0.3 & 0.5 & 0.15 & 1.00 & 0.09 & 1.00 & 0.05 & 1.00 & 0.05 & 1.00 \\ 
0.7 & -0.5 & 0.39 & 1.00 & 0.20 & 0.99 & 0.05 & 0.80 & 0.04 & 0.83 \\ 
0.7 & 0.0 & 0.37 & 1.00 & 0.20 & 1.00 & 0.05 & 0.93 & 0.05 & 0.94 \\ 
0.7 & 0.5 & 0.35 & 1.00 & 0.21 & 1.00 & 0.04 & 0.99 & 0.03 & 1.00 \\ \hline
\end{tabular}

\endgroup

\bigskip

(b) $\delta _{i}=1.0$

\begingroup \scalefont{0.8}\renewcommand{\arraystretch}{1.0}

\begin{tabular}{cc|cccc|cccc}
\hline
&  & \multicolumn{4}{|c}{Asymptotic} & \multicolumn{4}{|c}{Bootstrap} \\ 
\hline
$\alpha $ & $\rho $ & (1) & (2) & (3) & (4) & (1) & (2) & (3) & (4) \\ \hline
0.0 & -0.5 & 0.35 & 1.00 & 0.07 & 1.00 & 0.02 & 1.00 & 0.06 & 1.00 \\ 
0.0 & 0.0 & 0.30 & 1.00 & 0.09 & 1.00 & 0.05 & 1.00 & 0.02 & 1.00 \\ 
0.0 & 0.5 & 0.37 & 1.00 & 0.31 & 1.00 & 0.02 & 1.00 & 0.01 & 1.00 \\ 
0.3 & -0.5 & 0.35 & 1.00 & 0.11 & 1.00 & 0.03 & 1.00 & 0.05 & 1.00 \\ 
0.3 & 0.0 & 0.33 & 1.00 & 0.22 & 1.00 & 0.04 & 1.00 & 0.02 & 1.00 \\ 
0.3 & 0.5 & 0.32 & 1.00 & 0.25 & 1.00 & 0.03 & 1.00 & 0.02 & 1.00 \\ 
0.7 & -0.5 & 0.26 & 1.00 & 0.20 & 1.00 & 0.05 & 0.96 & 0.05 & 0.99 \\ 
0.7 & 0.0 & 0.28 & 1.00 & 0.25 & 1.00 & 0.06 & 1.00 & 0.04 & 1.00 \\ 
0.7 & 0.5 & 0.26 & 1.00 & 0.20 & 1.00 & 0.05 & 1.00 & 0.06 & 1.00 \\ \hline
\end{tabular}

\endgroup

\bigskip

(c) $\delta _{i}=1.5$

\begingroup \scalefont{0.8}\renewcommand{\arraystretch}{1.0}

\begin{tabular}{cc|cccc|cccc}
\hline
&  & \multicolumn{4}{|c|}{Asymptotic} & \multicolumn{4}{|c}{Bootstrap} \\ 
\hline
$\alpha $ & $\rho $ & (1) & (2) & (3) & (4) & (1) & (2) & (3) & (4) \\ \hline
0.0 & -0.5 & 0.12 & 1.00 & 0.12 & 1.00 & 0.06 & 1.00 & 0.05 & 1.00 \\ 
0.0 & 0.0 & 0.10 & 1.00 & 0.10 & 1.00 & 0.04 & 1.00 & 0.03 & 1.00 \\ 
0.0 & 0.5 & 0.12 & 1.00 & 0.11 & 1.00 & 0.07 & 1.00 & 0.06 & 1.00 \\ 
0.3 & -0.5 & 0.09 & 1.00 & 0.09 & 1.00 & 0.06 & 1.00 & 0.06 & 1.00 \\ 
0.3 & 0.0 & 0.09 & 1.00 & 0.09 & 1.00 & 0.04 & 1.00 & 0.05 & 1.00 \\ 
0.3 & 0.5 & 0.09 & 1.00 & 0.09 & 1.00 & 0.06 & 1.00 & 0.05 & 1.00 \\ 
0.7 & -0.5 & 0.06 & 1.00 & 0.06 & 1.00 & 0.03 & 0.95 & 0.04 & 1.00 \\ 
0.7 & 0.0 & 0.07 & 1.00 & 0.07 & 1.00 & 0.04 & 0.99 & 0.04 & 1.00 \\ 
0.7 & 0.5 & 0.07 & 1.00 & 0.07 & 1.00 & 0.04 & 1.00 & 0.05 & 1.00 \\ \hline
\end{tabular}

\endgroup

\bigskip

\parbox{16cm}
{\begin{small} Note: $T=100$. The true break dates are $T/2$. The null hypotheses are as follows: (1) $R=I$ and $r=(0.5,0.5)'$, (2) $R=I$ and $r=(0.525,0.475)'$, (3) $R=[1,-1]$ and $r=0$ and (4) $R=[1,-1]$ and $r=0.05$. Hence, the null rejection probabilities for (1) and (3) stand for the finite sample sizes while those for (2) and (4) are powers.
\end{small}}
\end{center}

\newpage

\begin{center}
Table 4. P-values of common break date tests, GLS-Wald/LR

\bigskip

(a) Bivariate systems, 1900-1992

\begingroup\scalefont{0.75}\renewcommand{\arraystretch}{1.0}

\begin{tabular}{cc|cc|cc|ccc}
\hline
&  & \multicolumn{2}{|c|}{Wald} & \multicolumn{2}{|c|}{LR} & 
\multicolumn{3}{|c}{Break Dates} \\ \hline
forc. & \multicolumn{1}{c|}{temp.} & asym. & boot. & asym. & boot. & forc. & 
temp. & comm. \\ \hline
\multicolumn{1}{l}{W} & \multicolumn{1}{l|}{G H} & - & 0.19 & 0.01 & 0.11 & 
1963 & 1985 & 1962 \\ 
\multicolumn{1}{l}{} & \multicolumn{1}{l|}{G N} & - & 0.36 & - & 0.21 & 1963
& 1976 & 1963 \\ 
\multicolumn{1}{l}{} & \multicolumn{1}{l|}{G B} & - & 0.18 & - & 0.08 & 1963
& 1941 & 1963 \\ 
\multicolumn{1}{l}{} & \multicolumn{1}{l|}{G K} & 0.01 & 0.52 & 0.01 & 0.30
& 1963 & 1976 & 1963 \\ 
\multicolumn{1}{l}{} & \multicolumn{1}{l|}{Avg G} & - & 0.21 & 0.01 & 0.22 & 
1963 & 1985 & 1963 \\ 
\multicolumn{1}{l}{} & \multicolumn{1}{l|}{N H} & - & 0.19 & - & 0.06 & 1963
& 1985 & 1962 \\ 
\multicolumn{1}{l}{} & \multicolumn{1}{l|}{N N} & - & 0.21 & - & 0.14 & 1962
& 1942 & 1962 \\ 
\multicolumn{1}{l}{} & \multicolumn{1}{l|}{Avg~N} & - & 0.31 & 0.01 & 0.15 & 
1963 & 1985 & 1962 \\ 
\multicolumn{1}{l}{} & \multicolumn{1}{l|}{S H} & - & 0.34 & 0.03 & 0.21 & 
1963 & 1976 & 1963 \\ 
\multicolumn{1}{l}{} & \multicolumn{1}{l|}{S N} & - & 0.26 & 0.20 & 0.65 & 
1963 & 1967 & 1963 \\ 
\multicolumn{1}{l}{} & \multicolumn{1}{l|}{Avg~S} & - & 0.03 & 0.03 & 0.25 & 
1963 & 1975 & 1963 \\ \hline
\multicolumn{1}{l}{TRF} & \multicolumn{1}{l|}{G H} & 0.05 & 0.51 & 0.01 & 
0.13 & 1966 & 1978 & 1966 \\ 
\multicolumn{1}{l}{} & \multicolumn{1}{l|}{G N} & 0.24 & 0.63 & 0.03 & 0.43
& 1966 & 1975 & 1966 \\ 
\multicolumn{1}{l}{} & \multicolumn{1}{l|}{G B} & 0.03 & 0.58 & - & 0.11 & 
1966 & 1942 & 1965 \\ 
\multicolumn{1}{l}{} & \multicolumn{1}{l|}{G K} & - & 0.04 & 0.04 & 0.51 & 
1965 & 1905 & 1966 \\ 
\multicolumn{1}{l}{} & \multicolumn{1}{l|}{Avg G} & 0.29 & 0.72 & 0.03 & 0.34
& 1966 & 1976 & 1966 \\ 
\multicolumn{1}{l}{} & \multicolumn{1}{l|}{N H} & - & 0.25 & 0.01 & 0.07 & 
1966 & 1985 & 1966 \\ 
\multicolumn{1}{l}{} & \multicolumn{1}{l|}{N N} & 0.09 & 0.63 & - & 0.22 & 
1965 & 1940 & 1965 \\ 
\multicolumn{1}{l}{} & \multicolumn{1}{l|}{Avg~N} & - & 0.38 & 0.01 & 0.20 & 
1966 & 1985 & 1965 \\ 
\multicolumn{1}{l}{} & \multicolumn{1}{l|}{S H} & 0.05 & 0.49 & 0.05 & 0.25
& 1966 & 1976 & 1966 \\ 
\multicolumn{1}{l}{} & \multicolumn{1}{l|}{S N} & 1.00 & 1.00 & 1.00 & 1.00
& 1966 & 1966 & 1966 \\ 
\multicolumn{1}{l}{} & \multicolumn{1}{l|}{Avg~S} & 0.05 & 0.39 & 0.29 & 0.62
& 1966 & 1975 & 1966 \\ \hline
W & \multicolumn{1}{l|}{\~{G} H} & 0.23 & 0.48 & 0.54 & 0.72 & 1963 & 1964 & 
1963 \\ 
& \multicolumn{1}{l|}{\~{G} N} & - & 0.01 & 0.25 & 0.50 & 1963 & 1957 & 1963
\\ 
& \multicolumn{1}{l|}{\~{G} B} & - & 0.15 & 0.34 & 0.67 & 1963 & 1956 & 1963
\\ 
& \multicolumn{1}{l|}{\~{G} K} & - & 0.04 & 0.17 & 0.45 & 1963 & 1956 & 1963
\\ 
& \multicolumn{1}{l|}{Avg \~{G}} & - & 0.09 & 0.40 & 0.62 & 1963 & 1960 & 
1963 \\ 
& \multicolumn{1}{l|}{\~{N} H} & 0.05 & 0.29 & 0.38 & 0.56 & 1963 & 1965 & 
1963 \\ 
& \multicolumn{1}{l|}{\~{N} N} & 0.01 & 0.29 & 0.54 & 0.77 & 1963 & 1965 & 
1963 \\ 
& \multicolumn{1}{l|}{Avg~\~{N}} & 0.02 & 0.28 & 0.43 & 0.66 & 1963 & 1965 & 
1963 \\ 
& \multicolumn{1}{l|}{\~{S} H} & - & 0.10 & 0.18 & 0.41 & 1963 & 1956 & 1963
\\ 
& \multicolumn{1}{l|}{\~{S} N} & - & - & 0.02 & 0.22 & 1963 & 1954 & 1963 \\ 
& \multicolumn{1}{l|}{Avg~\~{S}} & - & - & 0.05 & 0.22 & 1963 & 1955 & 1963
\\ \hline
TRF & \multicolumn{1}{l|}{\~{G} H} & 0.79 & 0.86 & 0.66 & 0.84 & 1966 & 1965
& 1966 \\ 
& \multicolumn{1}{l|}{\~{G} N} & 0.13 & 0.39 & 0.18 & 0.51 & 1966 & 1960 & 
1966 \\ 
& \multicolumn{1}{l|}{\~{G} B} & 0.33 & 0.61 & 0.34 & 0.66 & 1966 & 1960 & 
1966 \\ 
& \multicolumn{1}{l|}{\~{G} K} & 0.20 & 0.48 & 0.19 & 0.53 & 1966 & 1960 & 
1966 \\ 
& \multicolumn{1}{l|}{Avg \~{G}} & 0.15 & 0.41 & 0.34 & 0.64 & 1966 & 1960 & 
1966 \\ 
& \multicolumn{1}{l|}{\~{N} H} & 1.00 & 1.00 & 1.00 & 1.00 & 1966 & 1966 & 
1966 \\ 
& \multicolumn{1}{l|}{\~{N} N} & 0.83 & 0.89 & 0.69 & 0.87 & 1966 & 1965 & 
1966 \\ 
& \multicolumn{1}{l|}{Avg~\~{N}} & 0.80 & 0.86 & 0.83 & 0.91 & 1966 & 1965 & 
1966 \\ 
& \multicolumn{1}{l|}{\~{S} H} & 0.04 & 0.29 & 0.08 & 0.31 & 1966 & 1956 & 
1966 \\ 
& \multicolumn{1}{l|}{\~{S} N} & 0.24 & 0.49 & 0.04 & 0.31 & 1966 & 1960 & 
1966 \\ 
& \multicolumn{1}{l|}{Avg~\~{S}} & 0.03 & 0.23 & 0.05 & 0.26 & 1966 & 1956 & 
1966 \\ \hline
\end{tabular}

\endgroup

\vspace{0.2cm}

\parbox{16cm}
{\begin{small} Note: Each temperature series is denoted by two letters. For the first one, G, N and S stand for global, northern and southern hemispheric temperatures, respectively. For the second one, H, N, B and K stand for HadCRUT4, NASA, Berkeley and Karl. Avg stands for average. The filtered temperature series are denoted with a tilde. W and TRF stand for well-mixed green-house gases and total radiative forcing, respectively. Entries less than 0.01 are marked as "-".
\end{small}}
\end{center}

\newpage

\begin{center}
Table 4. P-values of common break date tests, GLS-Wald/LR (cont.)

\bigskip

(b) Bivariate systems, 1963-2014

\begingroup\scalefont{0.75}\renewcommand{\arraystretch}{1.0}

\begin{tabular}{cc|cc|cc|ccc}
\hline
&  & \multicolumn{2}{|c}{Wald} & \multicolumn{2}{|c}{LR} & 
\multicolumn{3}{|c}{Break Dates} \\ \hline
forc. & \multicolumn{1}{c|}{temp.} & asym. & boot. & asym. & boot. & forc. & 
temp. & comm. \\ \hline
W & \multicolumn{1}{l|}{G H} & 0.01 & 0.48 & 0.37 & 0.64 & 1992 & 2007 & 1992
\\ 
& \multicolumn{1}{l|}{G N} & 0.01 & 0.65 & 0.22 & 0.53 & 1992 & 2006 & 1992
\\ 
& \multicolumn{1}{l|}{G B} & - & 0.28 & 0.27 & 0.57 & 1992 & 1975 & 1992 \\ 
& \multicolumn{1}{l|}{G K} & 0.08 & 0.73 & 0.34 & 0.69 & 1992 & 2005 & 1992
\\ 
& \multicolumn{1}{l|}{Avg G} & 0.03 & 0.59 & 0.47 & 0.82 & 1992 & 2007 & 1992
\\ 
& \multicolumn{1}{l|}{N H} & 0.35 & 0.67 & 0.33 & 0.61 & 1992 & 1985 & 1992
\\ 
& \multicolumn{1}{l|}{N N} & 0.31 & 0.65 & 0.41 & 0.72 & 1992 & 1985 & 1992
\\ 
& \multicolumn{1}{l|}{Avg~N} & 0.33 & 0.66 & 0.35 & 0.65 & 1992 & 1985 & 1992
\\ 
& \multicolumn{1}{l|}{S H} & 0.08 & 0.70 & 0.27 & 0.62 & 1992 & 2005 & 1992
\\ 
& \multicolumn{1}{l|}{S N} & - & 0.09 & 0.01 & 0.06 & 1992 & 1972 & 1992 \\ 
& \multicolumn{1}{l|}{Avg~S} & - & 0.20 & 0.19 & 0.46 & 1992 & 1969 & 1992
\\ \hline
\multicolumn{1}{l}{TRF} & \multicolumn{1}{l|}{G H} & - & 0.25 & 0.11 & 0.35
& 1990 & 1971 & 1990 \\ 
\multicolumn{1}{l}{} & \multicolumn{1}{l|}{G N} & - & 0.34 & 0.10 & 0.28 & 
1991 & 2007 & 1991 \\ 
\multicolumn{1}{l}{} & \multicolumn{1}{l|}{G B} & - & 0.34 & 0.18 & 0.50 & 
1990 & 1976 & 1990 \\ 
\multicolumn{1}{l}{} & \multicolumn{1}{l|}{G K} & - & 0.01 & 0.18 & 0.43 & 
1991 & 1966 & 1991 \\ 
\multicolumn{1}{l}{} & \multicolumn{1}{l|}{Avg G} & - & 0.26 & 0.22 & 0.54 & 
1991 & 1971 & 1990 \\ 
\multicolumn{1}{l}{} & \multicolumn{1}{l|}{N H} & - & 0.14 & 0.09 & 0.29 & 
1990 & 1969 & 1990 \\ 
\multicolumn{1}{l}{} & \multicolumn{1}{l|}{N N} & - & 0.10 & 0.03 & 0.22 & 
1991 & 1972 & 1990 \\ 
\multicolumn{1}{l}{} & \multicolumn{1}{l|}{Avg~N} & - & 0.14 & 0.05 & 0.26 & 
1991 & 1971 & 1990 \\ 
\multicolumn{1}{l}{} & \multicolumn{1}{l|}{S H} & - & 0.33 & 0.23 & 0.55 & 
1991 & 2006 & 1990 \\ 
\multicolumn{1}{l}{} & \multicolumn{1}{l|}{S N} & - & 0.04 & 0.01 & 0.08 & 
1991 & 1980 & 1990 \\ 
\multicolumn{1}{l}{} & \multicolumn{1}{l|}{Avg~S} & - & 0.44 & 0.38 & 0.70 & 
1991 & 1983 & 1991 \\ \hline
W & \multicolumn{1}{l|}{\~{G} H} & 1.00 & 1.00 & 1.00 & 1.00 & 1992 & 1992 & 
1992 \\ 
& \multicolumn{1}{l|}{\~{G} N} & 0.78 & 0.88 & 0.46 & 0.68 & 1992 & 1991 & 
1992 \\ 
& \multicolumn{1}{l|}{\~{G} B} & 1.00 & 1.00 & 1.00 & 1.00 & 1992 & 1992 & 
1992 \\ 
& \multicolumn{1}{l|}{\~{G} K} & 0.77 & 0.87 & 0.55 & 0.75 & 1992 & 1991 & 
1992 \\ 
& \multicolumn{1}{l|}{Avg \~{G}} & 0.76 & 0.87 & 0.74 & 0.87 & 1992 & 1991 & 
1992 \\ 
& \multicolumn{1}{l|}{\~{N} H} & - & 0.28 & 0.17 & 0.37 & 1992 & 2007 & 1992
\\ 
& \multicolumn{1}{l|}{\~{N} N} & 0.82 & 0.91 & 0.88 & 0.94 & 1992 & 1991 & 
1992 \\ 
& \multicolumn{1}{l|}{Avg~\~{N}} & 1.00 & 1.00 & 1.00 & 1.00 & 1992 & 1992 & 
1992 \\ 
& \multicolumn{1}{l|}{\~{S} H} & 0.75 & 0.84 & 0.79 & 0.88 & 1992 & 1991 & 
1992 \\ 
& \multicolumn{1}{l|}{\~{S} N} & - & 0.14 & 0.06 & 0.19 & 1992 & 1984 & 1992
\\ 
& \multicolumn{1}{l|}{Avg~\~{S}} & 0.01 & 0.23 & 0.28 & 0.47 & 1992 & 1986 & 
1992 \\ \hline
TRF & \multicolumn{1}{l|}{\~{G} H} & 0.18 & 0.63 & 0.75 & 0.90 & 1990 & 1992
& 1991 \\ 
& \multicolumn{1}{l|}{\~{G} N} & 0.71 & 0.87 & 0.83 & 0.92 & 1990 & 1991 & 
1991 \\ 
& \multicolumn{1}{l|}{\~{G} B} & 0.55 & 0.83 & 0.82 & 0.94 & 1991 & 1992 & 
1991 \\ 
& \multicolumn{1}{l|}{\~{G} K} & 0.70 & 0.87 & 0.85 & 0.91 & 1990 & 1991 & 
1991 \\ 
& \multicolumn{1}{l|}{Avg \~{G}} & 0.58 & 0.82 & 0.95 & 0.94 & 1990 & 1991 & 
1991 \\ 
& \multicolumn{1}{l|}{\~{N} H} & - & 0.07 & 0.21 & 0.47 & 1990 & 2007 & 1991
\\ 
& \multicolumn{1}{l|}{\~{N} N} & 0.24 & 0.66 & 0.80 & 0.94 & 1990 & 1991 & 
1991 \\ 
& \multicolumn{1}{l|}{Avg~\~{N}} & 0.18 & 0.67 & 0.74 & 0.92 & 1990 & 1992 & 
1991 \\ 
& \multicolumn{1}{l|}{\~{S} H} & 0.48 & 0.78 & 0.90 & 0.94 & 1990 & 1991 & 
1991 \\ 
& \multicolumn{1}{l|}{\~{S} N} & - & 0.16 & 0.06 & 0.34 & 1990 & 1984 & 1990
\\ 
& \multicolumn{1}{l|}{Avg~\~{S}} & - & 0.26 & 0.34 & 0.64 & 1990 & 1986 & 
1990 \\ \hline
\end{tabular}

\endgroup

\vspace{0.2cm}

\parbox{16cm}
{\begin{small} Note: Each temperature series is denoted by two letters. For the first one, G, N and S stand for global, northern and southern hemispheric temperatures, respectively. For the second one, H, N, B and K stand for HadCRUT4, NASA, Berkeley and Karl. Avg stands for average. The filtered temperature series are denoted with a tilde. W and TRF stand for well-mixed green-house gases and total radiative forcing, respectively. Entries less than 0.01 are marked as "-".
\end{small}}
\end{center}

\newpage

\begin{center}
Table 5. P-values of common break date tests, OLS-Wald

\bigskip

(a) Bivariate systems, 1900-1992

\begingroup \scalefont{0.8}\renewcommand{\arraystretch}{1.0}

\begin{tabular}{lc|cc|cc}
\hline
&  & \multicolumn{2}{|c}{W (1963)} & \multicolumn{2}{|c}{TRF (1966)} \\ 
&  & asym. & boot. & asym. & boot. \\ \hline
G H & (1985) & - & 0.13 & - & 0.16 \\ 
G N & (1976) & - & 0.33 & 0.16 & 0.55 \\ 
G B & (1941) & - & 0.14 & 0.02 & 0.54 \\ 
G K & (1976) & 0.01 & 0.51 & 0.26 & 0.69 \\ 
Avg G & (1985) & - & 0.18 & - & 0.34 \\ 
N H & (1985) & - & 0.17 & - & 0.18 \\ 
N N & (1939) & - & 0.39 & 0.06 & 0.58 \\ 
Avg~N & (1985) & - & 0.28 & - & 0.31 \\ 
S H & (1976) & - & 0.33 & 0.05 & 0.47 \\ 
S N & (1966) & 0.01 & 0.32 & 1.00 & 1.00 \\ 
Avg~S & (1975) & - & 0.02 & 0.05 & 0.37 \\ \hline
\~{G} H & (1965) & 0.01 & 0.19 & 0.79 & 0.86 \\ 
\~{G} N & (1960) & - & - & 0.13 & 0.40 \\ 
\~{G} B & (1960) & - & 0.16 & 0.33 & 0.61 \\ 
\~{G} K & (1960) & - & 0.03 & 0.20 & 0.50 \\ 
Avg \~{G} & (1960) & - & 0.07 & 0.15 & 0.41 \\ 
\~{N} H & (1966) & - & 0.13 & 1.00 & 1.00 \\ 
\~{N} N & (1965) & 0.01 & 0.20 & 0.83 & 0.89 \\ 
Avg~\~{N} & (1965) & 0.02 & 0.22 & 0.80 & 0.86 \\ 
\~{S} H & (1956) & - & 0.06 & 0.04 & 0.34 \\ 
\~{S} N & (1960) & - & 0.15 & 0.25 & 0.53 \\ 
Avg~\~{S} & (1956) & - & - & 0.03 & 0.24 \\ \hline
\end{tabular}

\endgroup

\bigskip

\parbox{16cm}
{\begin{small} Note: Each temperature series is denoted by two letters. For the first one, G, N and S stand for global, northern and southern hemispheric temperatures, respectively. For the second one, H, N, B and K stand for HadCRUT4, NASA, Berkeley and Karl. Avg stands for average. The filtered temperature series are denoted with a tilde. W and TRF stand for well-mixed green-house gases and total radiative forcing, respectively. Entries less than 0.01 are marked as "-". The numbers in parentheses are break dates.
\end{small}}
\end{center}

\pagebreak

\begin{center}
Table 5. P-values of common break bate tests, OLS-Wald (cont.)

\bigskip

(b) Bivariate systems, 1963-2014

\begingroup \scalefont{0.8}\renewcommand{\arraystretch}{1.0}

\begin{tabular}{lc|cc|cc}
\hline
&  & \multicolumn{2}{|c}{W (1992)} & \multicolumn{2}{|c}{TRF (1990)} \\ 
&  & asym. & boot. & asym. & boot. \\ \hline
G H & (1974) & - & 0.29 & - & 0.26 \\ 
G N & (1966) & - & 0.16 & - & 0.01 \\ 
G B & (1975) & - & 0.26 & - & 0.21 \\ 
G K & (1966) & - & 0.10 & - & 0.00 \\ 
Avg G & (1974) & - & 0.38 & - & 0.24 \\ 
N H & (1972) & - & 0.15 & - & 0.18 \\ 
N N & (1972) & - & 0.12 & - & 0.07 \\ 
Avg~N & (1972) & - & 0.13 & - & 0.11 \\ 
S H & (2005) & 0.14 & 0.71 & - & 0.43 \\ 
S N & (1980) & - & 0.14 & - & 0.05 \\ 
Avg~S & (1987) & 0.44 & 0.75 & 0.12 & 0.61 \\ \hline
\~{G} H & (1992) & 1.00 & 1.00 & 0.18 & 0.64 \\ 
\~{G} N & (1991) & 0.78 & 0.89 & 0.71 & 0.88 \\ 
\~{G} B & (1992) & 1.00 & 1.00 & 0.28 & 0.68 \\ 
\~{G} K & (1991) & 0.77 & 0.89 & 0.70 & 0.89 \\ 
Avg \~{G} & (1991) & 0.77 & 0.87 & 0.59 & 0.83 \\ 
\~{N} H & (2007) & - & 0.21 & 0.00 & 0.07 \\ 
\~{N} N & (1991) & 0.82 & 0.89 & 0.25 & 0.67 \\ 
Avg~\~{N} & (1992) & 1.00 & 1.00 & 0.18 & 0.65 \\ 
\~{S} H & (1991) & 0.75 & 0.86 & 0.48 & 0.78 \\ 
\~{S} N & (1984) & - & 0.16 & - & 0.15 \\ 
Avg~\~{S} & (1985) & - & 0.15 & - & 0.14 \\ \hline
\end{tabular}

\endgroup

\bigskip

\parbox{16cm}
{\begin{small} Note: Each temperature series is denoted by two letters. For the first one, G, N and S stand for global, northern and southern hemispheric temperatures, respectively. For the second one, H, N, B and K stand for HadCRUT4, NASA, Berkeley and Karl. Avg stands for average. The filtered temperature series are denoted with a tilde. W and TRF stand for well-mixed green-house gases and total radiative forcing, respectively. Entries less than 0.01 are marked as "-". The numbers in parentheses are break dates.
\end{small}}
\end{center}

\newpage

\begin{center}
Table 6. P-values of common break date tests, OLS-Wald

\bigskip

(a) Multivariate systems, 1900-1992

\begingroup \scalefont{0.7}\renewcommand{\arraystretch}{1.0}

\begin{tabular}{ll|c|cc|c|cc|c|cc}
\hline
&  & \multicolumn{3}{|c|}{HadCRUT4} & \multicolumn{3}{|c|}{NASA} & 
\multicolumn{3}{|c}{Average} \\ \cline{3-11}
&  & p-value & \multicolumn{2}{|c|}{Break Dates} & p-value & 
\multicolumn{2}{|c|}{Break Dates} & p-value & \multicolumn{2}{|c}{Break Dates
} \\ \cline{3-11}
Forc. & Temp. & asymp./boot. & Forcing & Temp. & asymp./boot. & Forcing & 
Temp. & asymp./boot. & Forcing & Temp. \\ \hline
W & G,N,S & \ - \ \ / 0.22 & 1963 &  & \ - \ \ / 0.58 & 1963 &  & \ - \ \ /
0.10 & 1963 &  \\ 
& G & \ - \ \ / - \ \  &  & 1985 & \ - \ \ / 0.44 &  & 1976 & \ - \ \ / 0.26
&  & 1985 \\ 
& N & \ - \ \ / 0.53 &  & 1985 & \ - \ \ / 0.46 &  & 1939 & \ - \ \ / 0.30 & 
& 1985 \\ 
& S & \ - \ \ / 0.28 &  & 1976 & 0.13 / 0.52 &  & 1966 & \ - \ \ / 0.06 &  & 
1975 \\ 
TRF & G,N,S & \ - \ \ / 0.16 & 1966 &  & \ - \ \ / 0.57 & 1966 &  & \ - \ \
/ 0.13 & 1966 &  \\ 
& G & \ - \ \ / 0.43 &  & 1985 & 0.04 / 0.46 &  & 1976 & \ - \ \ / 0.41 &  & 
1985 \\ 
& N & \ -\ \ \ / 0.34 &  & 1985 & \ - \ \ / 0.41 &  & 1939 & \ - \ \ / 0.30
&  & 1985 \\ 
& S & 0.01 / 0.47 &  & 1976 & 1.00 / 0.95 &  & 1966 & \ - \ \ / 0.05 &  & 
1975 \\ \hline
W & \~{G},\~{N},\~{S} & \ - \ \ / 0.10 & 1963 &  & \ - \ \ / \ \ - \ \  & 
1963 &  & \ - \ \ / \ - \ \  & 1963 &  \\ 
& \~{G} & 0.06 / 0.35 &  & 1965 & \ - \ \ / \ \ - \ \  &  & 1960 & \ - \ \ /
0.12 &  & 1960 \\ 
& \~{N} & \ - \ \ / 0.34 &  & 1966 & 0.04 / 0.40 &  & 1965 & 0.05 / 0.32 & 
& 1965 \\ 
& \~{S} & \ - \ \ / 0.13 &  & 1956 & 0.02 / 0.48 &  & 1960 & \ - \ \ / \ - \
\  &  & 1956 \\ 
TRF & \~{G},\~{N},\~{S} & \ - \ \ / 0.08 & 1966 &  & \ - \ \ / \ \ - \ \  & 
1966 &  & \ - \ \ / \ - \ \  & 1966 &  \\ 
& \~{G} & 0.35 / 0.68 &  & 1965 & \ - \ \ / 0.03 &  & 1960 & \ - \ \ / 0.02
&  & 1960 \\ 
& \~{N} & 1.00 / 0.94 &  & 1966 & 0.17 / 0.57 &  & 1965 & 0.22 / 0.47 &  & 
1965 \\ 
& \~{S} & \ - \ \ / 0.11 &  & 1956 & \ - \ \ / 0.35 &  & 1960 & \ - \ \ / \
- \ \  &  & 1956 \\ \hline
\end{tabular}

\endgroup \bigskip

(b) Multivariate systems, 1963-2014

\begingroup \scalefont{0.7}\renewcommand{\arraystretch}{1.0}

\begin{tabular}{ll|c|cc|c|cc|c|cc}
\hline
&  & \multicolumn{3}{|c|}{HadCRUT4} & \multicolumn{3}{|c|}{NASA} & 
\multicolumn{3}{|c}{Average} \\ \cline{3-11}
&  & p-value & \multicolumn{2}{|c|}{Break Dates} & p-value & 
\multicolumn{2}{|c|}{Break Dates} & p-value & \multicolumn{2}{|c}{Break Dates
} \\ \cline{3-11}
Forc. & Temp. & asymp./boot. & Forcing & Temp. & asymp./boot. & Forcing & 
Temp. & asymp./boot. & Forcing & Temp. \\ \hline
W & G,N,S & \ - \ \ / 0.36 & 1992 &  & \ - \ \ / 0.10 & 1992 &  & \ - \ \ /
0.49 & 1992 &  \\ 
& G & \ - \ \ / 0.29 &  & 1974 & \ - \ \ / 0.07 &  & 1966 & \ - \ \ / 0.29 & 
& 1974 \\ 
& N & \ - \ \ / 0.19 &  & 1972 & \ - \ \ / 0.13 &  & 1972 & \ - \ \ / 0.16 & 
& 1972 \\ 
& S & 0.01/ 0.55 &  & 2005 & \ - \ \ / 0.13 &  & 1980 & 0.07 / 0.60 &  & 1987
\\ 
TRF & G,N,S & \ - \ \ / 0.36 & 1990 &  & \ - \ \ / 0.05 & 1990 &  & \ - \ \
/ 0.35 & 1990 &  \\ 
& G & \ - \ \ / 0.29 &  & 1974 & \ - \ \ / 0.01 &  & 1966 & \ - \ \ / 0.30 & 
& 1974 \\ 
& N & \ - \ \ / 0.18 &  & 1972 & \ - \ \ / 0.08 &  & 1972 & \ - \ \ / 0.12 & 
& 1972 \\ 
& S & \ - \ \ / 0.48 &  & 2005 & \ - \ \ / 0.05 &  & 1980 & 0.22 / 0.62 &  & 
1987 \\ \hline
W & \~{G},\~{N},\~{S} & \ - \ \ / 0.13 & 1992 &  & \ - \ \ / 0.21 & 1992 & 
& \ - \ \ / 0.15 & 1992 &  \\ 
& \~{G} & 1.00 / 0.93 &  & 1992 & 0.77 / 0.86 &  & 1991 & 0.70 / 0.82 &  & 
1991 \\ 
& \~{N} & \ - \ \ / 0.19 &  & 2007 & 0.48 / 0.78 &  & 1991 & 1.00 / 0.95 & 
& 1992 \\ 
& \~{S} & 0.63 / 0.77 &  & 1991 & \ - \ \ / 0.20 &  & 1984 & \ - \ \ / 0.14
&  & 1985 \\ 
TRF & \~{G},\~{N},\~{S} & \ - \ \ / 0.10 & 1990 &  & \ - \ \ / 0.18 & 1990 & 
& \ - \ \ / 0.14 & 1990 &  \\ 
& \~{G} & 0.23 / 0.57 &  & 1992 & 0.72 / 0.88 &  & 1991 & 0.61 / 0.83 &  & 
1991 \\ 
& \~{N} & \ - \ \ / 0.08 &  & 2007 & 0.27 / 0.71 &  & 1991 & 0.21 / 0.65 & 
& 1992 \\ 
& \~{S} & 0.50 / 0.70 &  & 1991 & \ - \ \ / 0.17 &  & 1984 & \ - \ \ / 0.13
&  & 1985 \\ \hline
\end{tabular}

\endgroup

\parbox{16cm}
{\begin{small} Note: G, N and S stand for global, northern and southern hemispheric temperatures, respectively. The filtered temperature series are denoted with a tilde. W and TRF stand for well-mixed green-house gases and total radiative forcing, respectively. Entries less than 0.01 are marked as "-".
  \end{small}}

\newpage

Table 7. P-values of joint common break date tests, OLS-Wald\vspace{0.2in}

- Bivariate systems, 1926-2014 -

\begingroup\scalefont{0.8}\renewcommand{\arraystretch}{1.0}

\begin{tabular}{lc|cc|cc}
\hline
&  & \multicolumn{2}{|c|}{W (1963, 1992)} & \multicolumn{2}{|c}{TRF (1965,
1990)} \\ \cline{3-6}
&  & asym. & boot. & asym. & boot. \\ \hline
G H & (1939, 1972) & - & 0.42 & - & 0.03 \\ 
G N & (1942, 1966) & - & 0.69 & - & 0.02 \\ 
G B & (1941, 1974) & - & 0.25 & - & 0.01 \\ 
G K & (1942, 1967) & - & 0.76 & - & 0.02 \\ 
Avg G & (1941, 1971) & - & 0.87 & - & 0.02 \\ 
N H & (1939, 1972) & - & 0.42 & - & 0.06 \\ 
N N & (1941, 1974) & - & 0.43 & - & 0.02 \\ 
Avg N & (1939, 1974) & - & 0.42 & - & 0.04 \\ 
S H & (1975, 1988) & - & 0.31 & - & 0.40 \\ 
S N & (1942, 1955) & - & 0.65 & - & 0.01 \\ 
Avg S & (1942, 1955) & - & 0.02 & - & 0.01 \\ \hline
\multicolumn{1}{l}{\~{G} H} & (1965, 1992) & - & 0.69 & 0.71 & 0.95 \\ 
\multicolumn{1}{l}{\~{G} N} & (1965, 1984) & - & 0.42 & 0.19 & 0.33 \\ 
\multicolumn{1}{l}{\~{G} B} & (1964, 1992) & 0.57 & 1.00 & 0.89 & 0.95 \\ 
\multicolumn{1}{l}{\~{G} K} & (1965, 1984) & - & 0.32 & 0.15 & 0.32 \\ 
\multicolumn{1}{l}{Avg \~{G}} & (1964, 1991) & 0.18 & 0.98 & 0.94 & 0.98 \\ 
\multicolumn{1}{l}{\~{N} H} & (1966, 1994) & - & 0.76 & 0.36 & 0.81 \\ 
\multicolumn{1}{l}{\~{N} N} & (1965, 1991) & - & 0.88 & 0.96 & 1.00 \\ 
\multicolumn{1}{l}{Avg~\~{N}} & (1966, 1991) & - & 0.69 & 0.83 & 0.97 \\ 
\multicolumn{1}{l}{\~{S} H} & (1955, 1991) & - & 0.13 & - & 0.08 \\ 
\multicolumn{1}{l}{\~{S} N} & (1965, 1984) & - & 0.49 & 0.20 & 0.26 \\ 
\multicolumn{1}{l}{Avg~\~{S}} & (1964, 1985) & - & 0.28 & 0.05 & 0.23 \\ 
\hline
\end{tabular}

\endgroup

\bigskip

\parbox{16cm}
{\begin{small} Note: G, N and S stand for global, northern and southern hemispheric temperatures, respectively. The filtered temperature series are denoted with a tilde. W and TRF stand for well-mixed green-house gases and total radiative forcing, respectively. Entries less than 0.01 are marked as "-".
\end{small}}

\newpage

Table 8: P-values of the LR test for no break in temperature series

Filtered temperature series and anthropogenic forcing

\begingroup\scalefont{0.7}\renewcommand{\arraystretch}{1.0}

\begin{tabular}{llccccccc}
\hline
&  & \multicolumn{3}{c}{1900-1992 ($T=93$)} &  & \multicolumn{3}{c}{
1963-2014 ($T=52$)} \\ \cline{3-5}\cline{7-9}
&  & $p$-value &  & Break date &  & $p$-value &  & Break date \\ 
Forc. & Temp. & ($LR$ test) &  & under $H_{1}$ &  & ($LR$ test) &  & under $%
H_{1}$ \\ \hline
W & $\tilde{G}$ H & 0.00 &  & 1964 &  & 0.08 &  & 1992 \\ 
& $\tilde{G}$ N & 0.00 &  & 1957 &  & 0.01 &  & 1991 \\ 
& $\tilde{G}$ B & 0.04 &  & 1956 &  & 0.12 &  & 1992 \\ 
& $\tilde{G}$ K & 0.00 &  & 1956 &  & 0.02 &  & 1991 \\ 
& Avg $\tilde{G}$ & 0.01 &  & 1960 &  & 0.04 &  & 1991 \\ 
& $\tilde{N}$ H & 0.01 &  & 1965 &  & 0.16 &  & 2007 \\ 
& $\tilde{N}$ N & 0.01 &  & 1965 &  & 0.19 &  & 1991 \\ 
& Avg $\tilde{N}$ & 0.00 &  & 1965 &  & 0.21 &  & 1992 \\ 
& $\tilde{S}$ H & 0.00 &  & 1956 &  & 0.03 &  & 1991 \\ 
& $\tilde{S}$ N & 0.01 &  & 1954 &  & 0.00 &  & 1984 \\ 
& Avg $\tilde{S}$ & 0.00 &  & 1955 &  & 0.01 &  & 1986 \\ \hline
TRF & $\tilde{G}$ H & 0.00 &  & 1965 &  & 0.06 &  & 1992 \\ 
& $\tilde{G}$ N & 0.00 &  & 1960 &  & 0.01 &  & 1991 \\ 
& $\tilde{G}$ B & 0.04 &  & 1960 &  & 0.09 &  & 1992 \\ 
& $\tilde{G}$ K & 0.01 &  & 1960 &  & 0.02 &  & 1991 \\ 
& Avg $\tilde{G}$ & 0.00 &  & 1960 &  & 0.03 &  & 1991 \\ 
& $\tilde{N}$ H & 0.00 &  & 1966 &  & 0.15 &  & 2007 \\ 
& $\tilde{N}$ N & 0.01 &  & 1965 &  & 0.14 &  & 1991 \\ 
& Avg $\tilde{N}$ & 0.00 &  & 1965 &  & 0.19 &  & 1992 \\ 
& $\tilde{S}$ H & 0.00 &  & 1956 &  & 0.02 &  & 1991 \\ 
& $\tilde{S}$ N & 0.00 &  & 1960 &  & 0.00 &  & 1984 \\ 
& Avg $\tilde{S}$ & 0.00 &  & 1956 &  & 0.01 &  & 1986 \\ \hline
\end{tabular}

\endgroup

\parbox{16cm}
{\begin{small} Note: Each temperature series is denoted by two letters. For the first one, G, N and S stand for global, northern and southern hemispheric temperature respectively. For the second one, H, N, B and K stand for HadCRUT4, NASA, Berkeley and Karl. Avg stands for average. W and TRF stand for well-mixed green-house gases and total radiative forcing, repectively. The third and fifth columns report p-values for the LR test obtained from a bootstrap procedure and the fourth and sixth columns present the break date estimate in the temperature series under the alternative hypothesis given an estimated break in the forcing series. 
\end{small}}
\end{center}

\newpage

\setcounter{page}{1}\renewcommand{\thepage}{F-\arabic{page}}

\begin{figure}[h]
  \centering
  \includegraphics[width=1.0\textwidth]{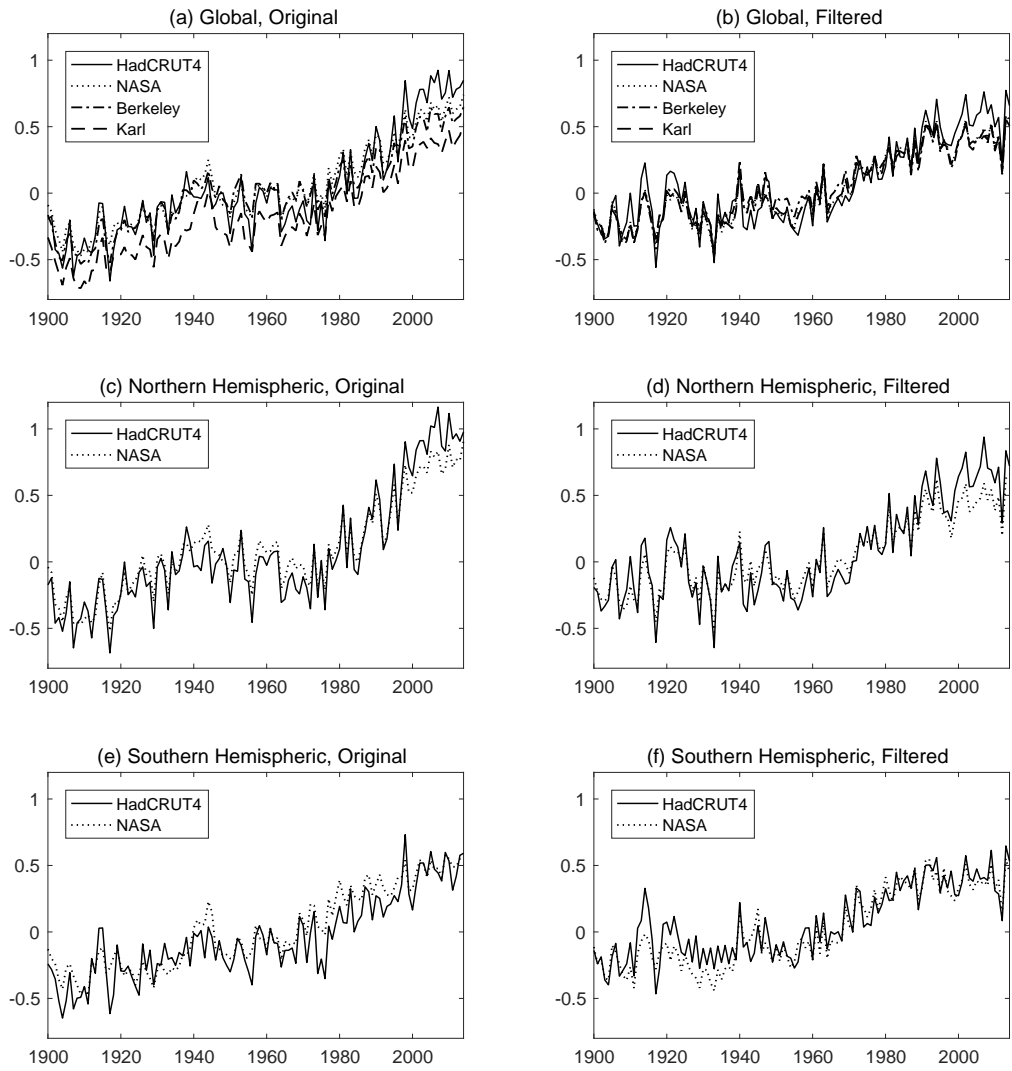}
  \caption{Global, Northern and Southern Hemispheric Temperature Series}    
\end{figure}

\newpage

\begin{figure}
  \centering
  \includegraphics[width=1.0\textwidth]{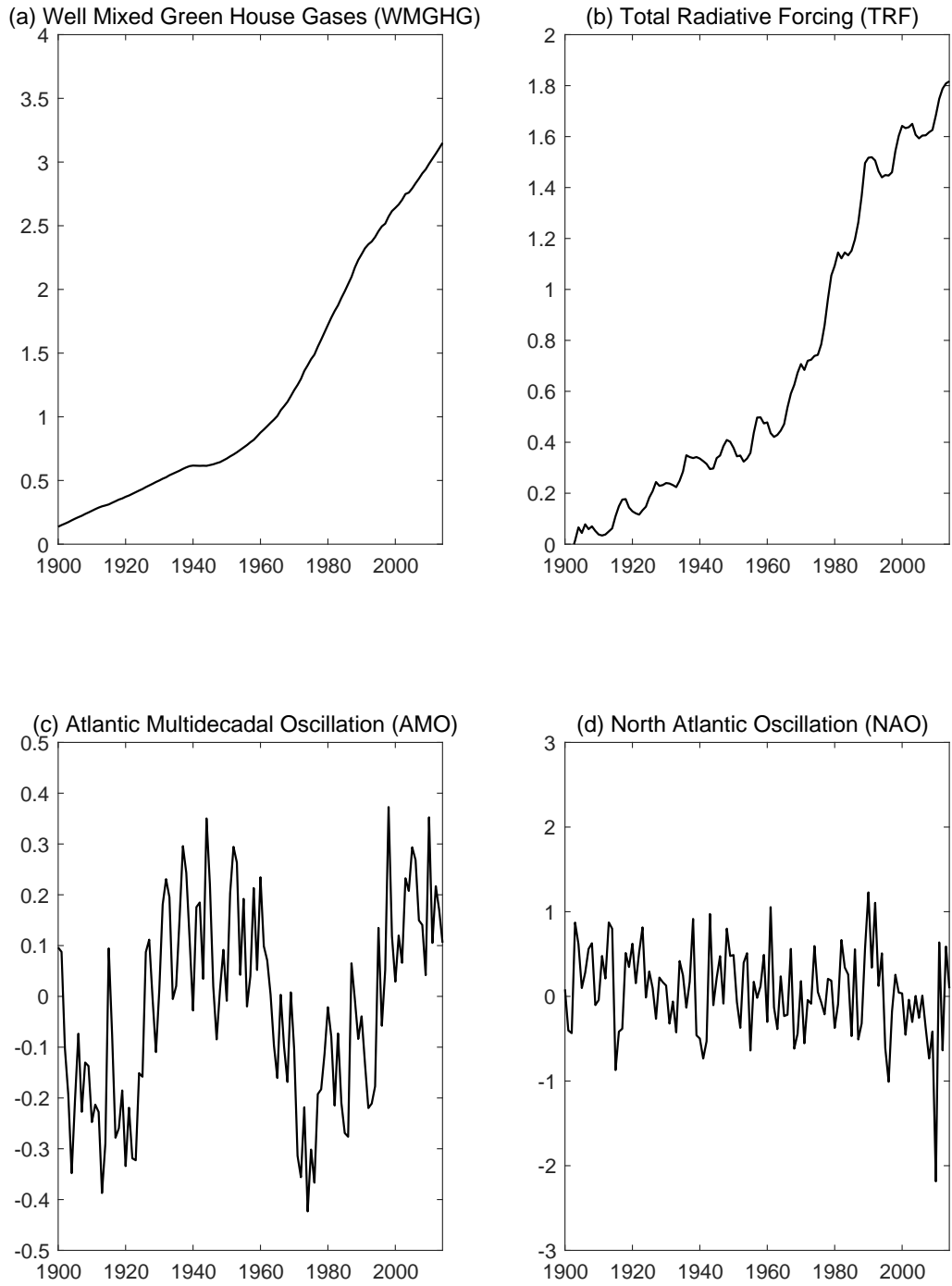}
  \caption{Radiative Forcing Variables and Modes of Variablility}
\end{figure}

\end{document}